\documentclass[12pt]{iopart}
\usepackage[pdftex]{graphicx}
\usepackage{axodraw}
\usepackage{amssymb}

\begin{document}

\title{Leptogenesis: beyond the minimal type I seesaw
scenario}

\author{Thomas Hambye}

\address{Service de Physique Th\'eorique,
Universit\'e Libre de Bruxelles, 1050 Brussels, Belgium\\
\vspace{0mm}
Departamento de F\'isica Te\'orica, Universidad Aut\'onoma de Madrid and
Instituto de F\'isica Te\'orica IFT-UAM/CSIC, Cantoblanco, 28049 Madrid, Spain}
\ead{thambye@ulb.ac.be}
\begin{abstract}
Numerous recent evidences for neutrino masses have established the leptogenesis mechanism as a very natural possible explanation for the baryon asymmetry of the Universe. The explicit realization of this mechanism depends on the neutrino mass model considered. If the right-handed type-I seesaw model of neutrino masses is certainly the most straightforward, it is not the only natural one, especially in the framework of explicit GUT realizations of the seesaw. In this review we discuss in detail the various seesaw scenarios that can implement the leptogenesis mechanism successfully, beyond the paradigm of the pure standard type-I seesaw model. This includes scenarios based on the existence of scalar triplets (type-II), of fermion triplets (type-III) as well as mixed seesaw frameworks. 
\end{abstract}

\maketitle

\section{Introduction}

The origin of the baryon asymmetry of the Universe constitutes one of the most fascinating enigma of contemporary physics. It definitely deserves an explanation! Such an explanation has not been found within the Standard Model (SM) of elementary interactions. But nothing prevents that it could be found beyond the SM. Regardless of the baryogenesis enigma, there are many reasons to believe that the SM does not constitute a complete theory anyway. In particular the many recent evidences for the neutrino masses (the clearest laboratory evidence one has at the moment for new physics) require the existence of new states beyond the SM, still to be found experimentally.
Especially attractive are neutrino mass models where these masses are induced by seesaw states, right-handed neutrinos ("type-I"), scalar triplets ("type-II") or fermion triplets ("type-III").
It is a quite intriguing fact that in most models of neutrino masses, and especially seesaw models, the interactions at the origin of neutrino masses do induce a baryon asymmetry in the Universe, through the leptogenesis mechanism \cite{Fukugita:1986hr}. With the exception of one scenario, the one that involves only a single scalar triplet, see below, this is basically unavoidable, unless one adopts specific assumptions and symmetries which restrict the interactions of these new states,
or unless one adopts specific cosmological scenarios (e.g.~a low reheating temperature at the end of the inflation era, so that the heavy states at the origin of the neutrino masses have never been produced in the thermal bath of the Universe). In particular in most neutrino mass models the interactions at the origin of the neutrino masses generically do satisfy the 3 Sakharov conditions for creating a baryon asymmetry. They break lepton number (hence baryon number together with the SM "sphalerons"), they break C and CP and are generally expected to be out-of-equilibrium during various epochs of the thermal history of the Universe. Moreover, even if there is unfortunately no one-to-one connection between the size of the neutrino masses and the baryon asymmetry produced, see below, and even if the model at the origin of the neutrino masses could easily lead to a too small baryon asymmetry, it is also intriguing that the size of the neutrino masses is in the ballpark of the values we need to induce easily a large enough baryon asymmetry. 
This constitutes a too good opportunity not to study it in details. For the type-I seesaw model this has been reviewed at length in quite a few references, see e.g.~Ref.~\cite{blanchet} in this volume and Refs.~\cite{Buchmuller:2004nz,Strumia:2006qk,Davidson:2008bu,Giudice:2003jh,Hambye:2003rt,Hambye:2004fn}. This review is dedicated to leptogenesis in the framework of the seesaw models beyond the minimal type-I seesaw framework, that is to say in the type-II and type-III models, as well as in mixed seesaw scenarios, type-I + type-II, type-I + type-III, etc. 

In Section 2 we begin by introducing and motivating the various seesaw models. Section 3 introduces a few model independent general considerations on the CP asymmetries. Section 4 (5) discuss in details how leptogenesis comes into play from the decay of a type-III (type-II) seesaw state, giving in particular many explanations on how the "efficiency factor" behaves in these frameworks. Many comparisons are done between these models and with the type-I seesaw model. Finally, Section 6 considers mixed seesaw scenarios, in particular the ones which come naturally out of minimal Grand Unified Theory (GUT) models.

\section{The various seesaw models and their motivation}

The most straighforward and, in many respects, the most  "natural" mechanism one can conceive to generate naturally small neutrino masses is the well-known seesaw mechanism. This mechanism lies their origin in the existence of a L-violating dimension-5 Weinberg operator interaction
\begin{equation}
{\cal L}_{eff}=\frac{c^{d=5}_{\alpha\beta}}{\Lambda} (\overline{L^c}_\alpha \tilde{H}^\ast)(\tilde{H}^\dagger L_\beta)\,,
\end{equation}
 with $L=(\nu_\alpha,l^-_\alpha)^T$, $H=(H^+,H^0)^T$, $\tilde{H}=i\tau_2 H^\ast$ and $\Lambda$ an energy scale. This operator generates naturally small Majorana neutrino masses, in accordance with data, if the $\Lambda$ scale is large, 
 \begin{equation}
 {\cal M}_{\nu_{\alpha\beta}}=-\frac{v^2}{2}c^{d=5}_{\alpha\beta}/\Lambda\,,
 \end{equation}
with $v=246$~GeV. The Weinberg interaction is the only gauge invariant dimension 5 interaction one can write out of SM fields. Therefore prior to the discovery of the neutrino masses, one could have expected that its effect, that is to say the neutrino masses, would be the first laboratory manifestation one would observe of new physics at a high scale (far above the electroweak scale). This turns out to have happened (even if of course this discovery doesn't necessarily imply that the scale $\Lambda$ is far beyond the electroweak scale, since the coefficients $c^{d=5}_{\alpha\beta}$ also come into play).

\begin{figure}[t]
\centering
\includegraphics[height=3.0cm]{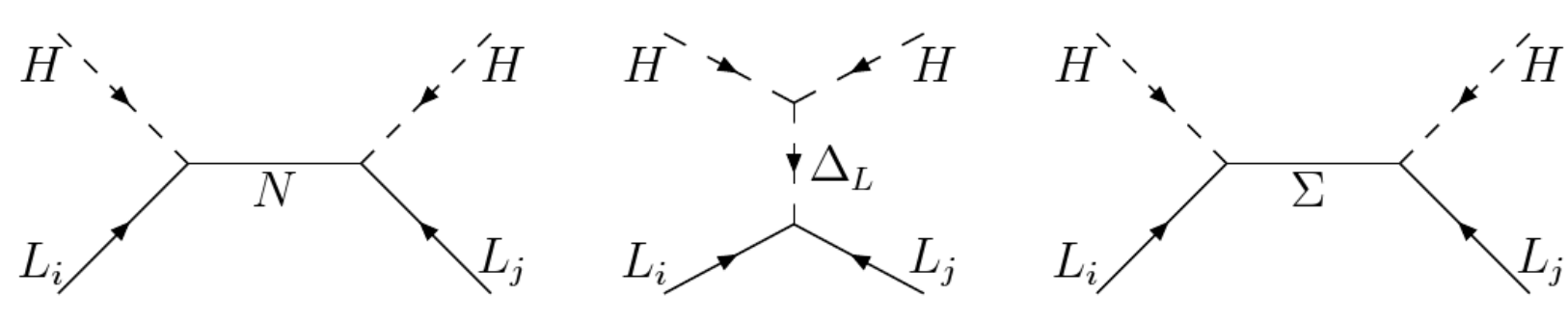}\\
\caption{The 3 basic seesaw diagrams which can induce naturally small neutrino masses.}
\label{fig1}
\end{figure}

There is in principle an infinite number of new possible states which could induce such an operator with large scale $\Lambda$. Nevertheless there are only three basic ways to generate it at tree level, the most straightforward way: from the exchange of one or several right-handed neutrinos, scalar triplets or fermion triplets, see Fig.~\ref{fig1}.\footnote{This can be understood easily from the fact that both $L$ and $H$ are doublets, and with 2 doublets coupling to the heavy states one can do only singlets or triplets of $SU(2)_L$: a right-handed neutrino \cite{seesaw}-\cite{seesaw5} (i.e.~$LH$ singlet), a scalar triplet \cite{typeIIseesaw}-\cite{typeIIseesaw4} (a $LL$ or $HH$ triplet), or a fermion triplet \cite{typeIIIseesaw} (a $LH$ triplet). As for the scalar singlet combination it doesn't give any neutrino masses because the $HH$ antisymmetric singlet contribution vanishes.} 
The relevant Lagrangian for each seesaw model is given in Table. 1, together with the neutrino mass matrix  it gives. In the fermionic cases the lepton number violation comes from the coexistence of Majorana masses and Yukawa couplings, whereas for the scalar triplet option it comes from the coexistence of the coupling to two Higgs doublets (L=0) and 2 lepton doublets (L=2). In all cases the masses of the new states are not protected by any SM symmetry, hence could be naturally much larger than the electroweak scale, leading to a large $\Lambda$ effective scale, and hence to naturally small neutrino masses, as observed.
The seesaw states with such large mass scales are typically expected in the GUT
 framework, another motivation for the seesaw framework.

%
\vspace{2cm}
\begin{table}[!h]
\begin{center}
\begin{tabular}{|c||c|c|c|} 
\hline
& \multicolumn{3}{c|}{Type of seesaw model} \rule[-5 pt]{0pt}{18 pt}\\
\cline{2-4}
 & {Type-I} & Type-II & Type-III  \rule[-5 pt]{0pt}{18 pt}\\
\hline
\hline
 {\scriptsize {Seesaw states} }& {\scriptsize $N$} & {\scriptsize $\Delta_L=
\left(
\begin{array}{ cc}
   \delta^+/\sqrt{2}  &   \delta^{++} \\
     \delta^0 &  -\delta^+/\sqrt{2} 
\end{array}
\right)$ }& {\scriptsize $\Sigma=
\left(
\begin{array}{ cc}
   \Sigma^0/\sqrt{2}  &   \Sigma^+ \\
     \Sigma^- &  -\Sigma^0/\sqrt{2} 
\end{array}
\right)$ }  \rule[-18 pt]{0pt}{44 pt} \\ 
\hline
 {\scriptsize {Kin. term} }& {\scriptsize $i\overline{N}{\partial\hspace{-4pt}\slash} N$ }& {\scriptsize $Tr[(D_\mu \Delta_L)^\dagger (D^\mu \Delta_L]$ } & {\scriptsize $Tr [ \overline{\Sigma} i \slash \hspace{-2.5mm} D  \Sigma ] $ }\rule[-14 pt]{0pt}{34 pt}\\
 \hline
 {\scriptsize {Mass term}} & {\scriptsize $-\frac{1}{2} Tr [\overline{N}  m_N N^c 
                +\overline{N^c} m_N^* N] $ }& {\scriptsize $-m^2_\Delta Tr[\Delta_L^\dagger \Delta_L]$} & {\scriptsize $-\frac{1}{2} Tr [\overline{\Sigma}  m_\Sigma \Sigma^c 
                +\overline{\Sigma^c} m_\Sigma^* \Sigma] $ } \rule[-14pt]{0pt}{34pt}\\
                 \hline
 {\scriptsize {Interactions}} & {\scriptsize $ - \tilde{\phi}^\dagger \overline{N} Y_N L 
-  \overline{L} {Y_N}^\dagger  N \tilde{\phi}$ } & {\scriptsize $-L^TY_\Delta C i \tau_2\Delta_LL+\mu \tilde{H}^Ti \tau_2 \Delta_L \tilde{H}$ } & {\scriptsize $- \tilde{\phi}^\dagger \overline{\Sigma} \sqrt{2}Y_\Sigma L 
-  \overline{L}\sqrt{2} {Y_\Sigma}^\dagger  \Sigma \tilde{\phi} $ } \rule[-14 pt]{0pt}{34 pt}\\
                 \hline
    {\scriptsize  {$\nu$ masses}} & {\scriptsize $ {\cal M}_\nu^N=-\frac{v^2}{2} Y_N^T\frac{1}{m_N} Y_N$ } & {\scriptsize $ {\cal M}_\nu^\Delta=2Y_\Delta v_{\Delta_L}=Y_\Delta\mu^*\frac{v^2}{m^2_\Delta}$ } & {\scriptsize $ {\cal M}_\nu^\Sigma=-\frac{v^2}{2} Y_\Sigma^T\frac{1}{m_\Sigma} Y_\Sigma$ } \rule[-14 pt]{0pt}{34 pt}\\
                 \hline
 {\scriptsize CP asym.} & {\scriptsize $ \varepsilon_N\equiv\frac{\Gamma(N\rightarrow L H)-\Gamma(N\rightarrow\overline{L}\bar{H})}{\Gamma(N\rightarrow L H)+\Gamma(N\rightarrow\overline{L}\bar{H})}$ }& {\scriptsize $\varepsilon_\Delta\equiv2\, \frac{\Gamma(\bar{\Delta}_L\rightarrow L L)-\Gamma(\Delta_L\rightarrow\overline{L}\overline{L})}{\Gamma_\Delta+\Gamma_{\bar{\Delta}}}$ }& {\scriptsize $\varepsilon_\Sigma\equiv \frac{\Gamma(\Sigma \rightarrow L H)-\Gamma(\overline{\Sigma}\rightarrow\overline{L}\bar{H})}{\Gamma(\Sigma\rightarrow L H)+\Gamma(\overline{\Sigma} \rightarrow\overline{L}\bar{H})}$}

\rule[-14 pt]{0pt}{34 pt}\\
\hline

\end{tabular}
\end{center}
\caption{\normalsize Lagrangian, neutrino mass contribution and CP-asymmetry definition for each seesaw model (in matrix notation, a summation over the lepton flavour and heavy state indices is implicit).
We define $\Sigma^c$ as $(\Sigma^c)_{ij}=(\Sigma_{ji})^c$ with $\psi^c\equiv C\overline{\psi}^T$ and $i,j=1,2$ the $SU(2)_L$ indices. For the scalar triplet $\Gamma_\Delta=\Gamma_{\bar{\Delta}}=\Gamma(\Delta_L\rightarrow\bar{L}\bar{L}) +\Gamma(\Delta_L\rightarrow {H} {H})$.}
\label{tableseesaw}
\end{table}
\normalsize

Among the various possibilities, the right-handed neutrino option (type-I) is the most considered, probably motivated by the intuition that the SM left-handed neutrinos, as all other fermions in the SM, should have right-handed partners. This is furthermore supported by the usual left-right \cite{leftright}-\cite{leftright4} and SO(10) GUT frameworks where right-handed neutrinos come automatically in, being associated to the SM fermions within a same representation (in particular in the $16_F$ representation of SO(10) together with a full generation of SM fermions). This is nevertheless not the only natural option, and even not necessarily the most natural one in these models.
One issue is that, if one assumes right-handed neutrinos with Majorana masses, these masses constitute new scales whose origin and values must be explained.
In these frameworks such scales are protected by the gauge symmetries, and therefore find their origin in the breaking of these gauge symmetries (for example of~$U(1)_{B-L}$ or $SU(2)_R$ which are contained in $SO(10)$). In the SO(10) context the right-handed neutrinos masses can be induced in a renormalizable way (if one doesn't want to induce them from higher dimensional operators, which would require the introduction of yet a new mass scale, beyond the GUT scale, or would need to rely on unknown Planck scale effects). This can be done only by introducing a 126 scalar representation.\footnote{As well known, as long as one assumes the right-handed neutrinos in the $16_F$ their masses can come only from the scalar 126 representation in the product 16$_F \times 16_F=126+120+10$.} This scalar representation has the particularity of containing both a right-handed scalar triplet $\Delta_R$, whose vev gives a mass to the right-handed neutrino, and a left-handed triplet $\Delta_L$, whose vev induces left-handed neutrino masses through the type-II seesaw mechanism. From the point of view of SO(10) a mixed type-I+type-II neutrino mass model is consequently a  natural option.
Similarly, from the point of view of SU(5), right-handed neutrinos are not especially more natural than any other extra fermions, since they are difficult to associate to any SM fermion within a same SU(5) representation.
One could introduce one or several of them as singlet representation of SU(5) but they could come as well from an adjoint representation of SU(5), for example. The 24 representation of SU(5) has the particularity to involve one fermion with the SM quantum numbers of a right-handed neutrino, but also to involve a fermionic triplet as in the type-III model. This framework offers therefore the opportunity to generate the neutrino masses from a single representation (like the type-II seesaw with only one scalar triplet). In other words SU(5) doesn't point less towards a type-I+type-III seesaw generation of neutrino masses than it does towards a pure type-I framework.

In the next 3 sections we consider the possibility of inducing successful leptogenesis in the framework of the type-II and type-III seesaw models, as well as in seesaw frameworks involving several types of seesaw. For each framework we will explicitate how the 2 main ingredients of leptogenesis come into play, the CP asymmetry and the efficiency factor. 

%

\section{A few model independent considerations on the CP asymmetries}

Along the leptogenesis scenario of baryogenesis, a lepton asymmetry is created in the Universe when the heavy seesaw states decay at a temperature of order their masses. The CP asymmetry is nothing but the average $\Delta L$ created each time a heavy state decays. This is simply given by the difference of the decay width to leptons and antileptons of these states, divided per their total decay width, multiplied by the number of lepton created in the decay of a single heavy state, see Table.1. Note that we will start by discussing the CP asymmetry for the total lepton number,
rather than the asymmetry for each flavour lepton number separately.
That is to say we will not take into account the possible effect of the Yukawa interactions of the charged leptons, which differentiate these equations. For seesaw states masses below $10^{12-13}$~GeV this may induce both conceptual and quantitative differences. Various flavour effects and associated references will be discussed in Sections~4.3 and 5.5. We begin by listing a number of important properties of the CP-asymmetries (some of them remain valid in the flavour case too, see below):
\begin{itemize}
\item(i) At lowest order, CP-violation can manifest itself only at the level of the interference between the tree level and  one-loop decay width contributions. The various one-loop diagrams which could contribute in the three seesaw framework as well as in various mixed seesaw frameworks are given in Fig.~\ref{fig2} and Fig.~\ref{fig33}. As well-known, the CP-violating imaginary part of the couplings must be associated to the absorptive part of the loop diagram. The sum of the masses of the particles involved in the loop must consequently be smaller than the mass of the decaying state.
\item(ii) CP violation requires moreover (at least for the unflavoured case) that the lepton number violation be explicit in the one loop diagram, i.e.~the diagrams must involve intermediate and final states with different lepton numbers. For example, as shown in Fig.~\ref{fig33}, in the type-II seesaw model one has to go through a HH intermediate state before going to the 2 lepton final state. In the type-I or type-III model one has an interference between a $LH$ state and a $\bar{L}H^*$ state, Fig.~\ref{fig2}.
 \begin{figure}[t]
\centering
\includegraphics[height=3.0cm]{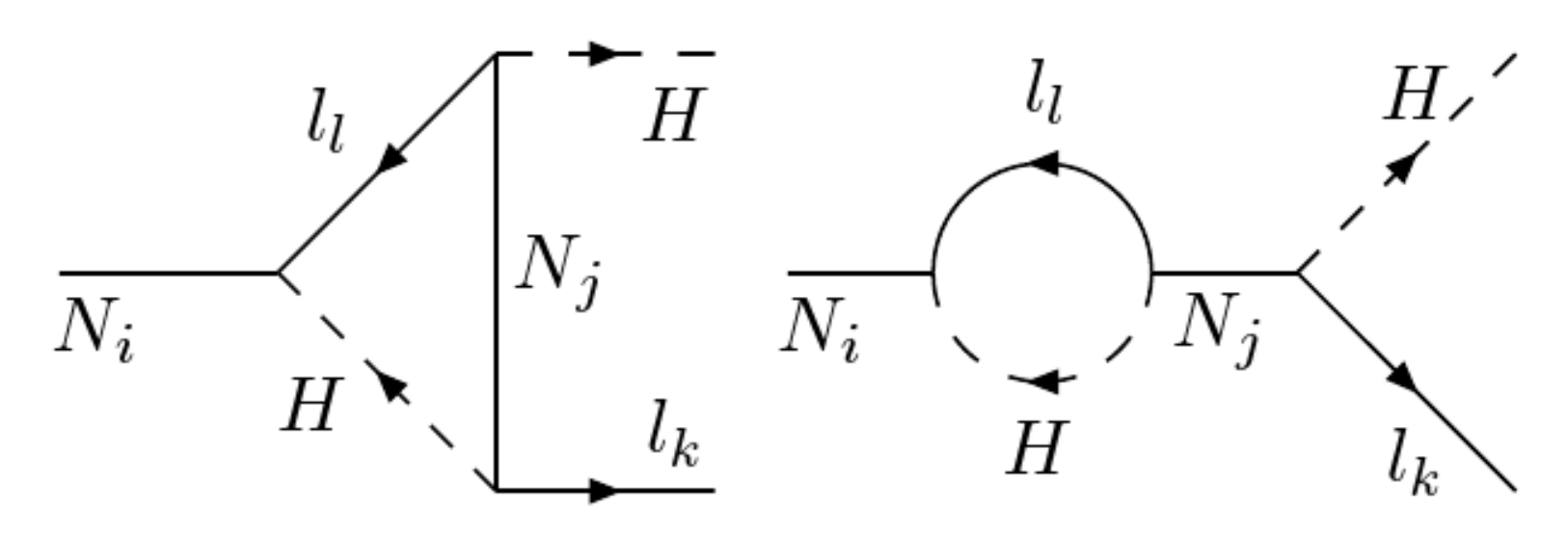}\\
\caption{One-loop diagrams contributing to the CP-asymmetry
 in the type-I case or type-III case (replacing $N_{i,j}$ by $\Sigma_{i,j}$).}
\label{fig2}
\end{figure}
\item(iii) A one-loop non zero asymmetry requires flavour "breaking" or more exactly at least two sources of flavour "breaking", i.e.~two heavy states with unequal couplings to leptons and/or scalar bosons. If there is only one heavy state or if there are several ones with exactly the same couplings to leptons and scalar bosons, the CP-asymmetry vanishes. In this case each coupling to lepton is automatically accompanied by its complex conjugate in the CP asymmetry. This is the reason why the pure type-II seesaw model with a single scalar triplet, which is the only one which can give 2 or 3 light neutrino masses (as required experimentally) from a single heavy state, gives a vanishing asymmetry, even though it a priori satisfies all Sakharov conditions, see Fig.~\ref{fig33} and Section 5.3.
\item(iv) Both heavy states must be non-degenerate in mass. Otherwise the sum over all intermediate and final flavours gives a real combination of couplings leading to a vanishing asymmetry.
 \begin{figure}[t]
\centering
\includegraphics[height=3.0cm]{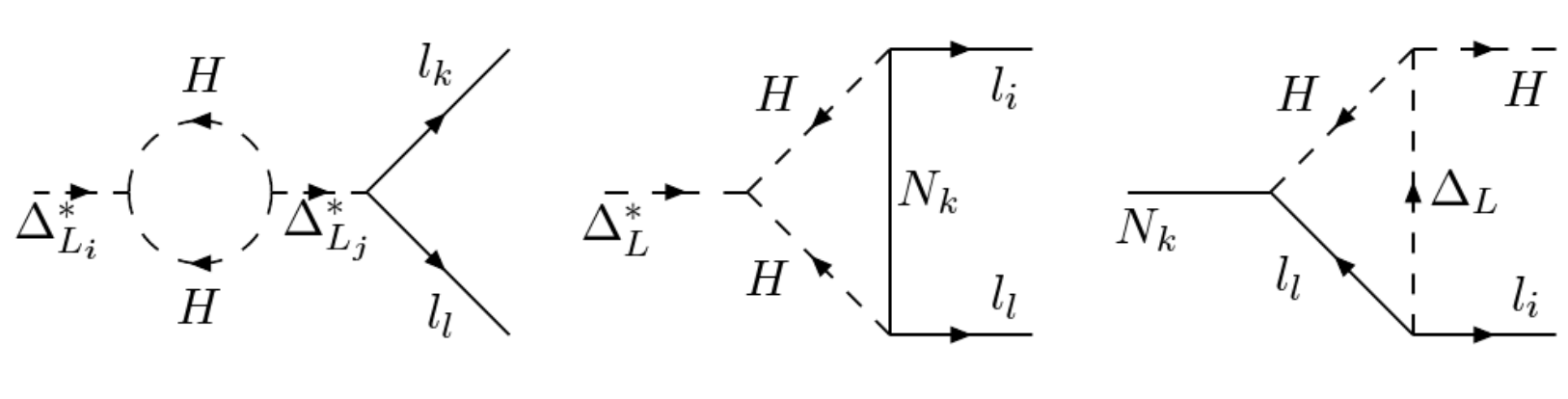}\\
\caption{Diagram contributing to the asymmetry
created by the decay of a $\Delta_L$ if there are several scalar triplets or a $N$. The third diagram also contribute to the asymmetry of a $N$ in presence of a heavier scalar triplet.}
\label{fig33}
\end{figure}
\item(v) In the limit where the virtual seesaw state mass is much larger than the one of the decaying state, i.e.~if the heavy states have a hierarchical mass spectrum (as quarks and leptons in the SM), the contribution of this virtual state can be fully parameterized by its contribution to the neutrino mass matrix (neglecting the charged lepton mass effects). The CP-asymmetry is proportional to it. This stems from the simple argument that the CP asymmetry comes from the absorptive part of the loop diagram, where particles inside the loop are on-shell. Therefore, to calculate the absorptive part, the heavy intermediate state can be integrated out, even though there is a loop diagram.
As a result the CP asymmetry is necessarily proportional to the neutrino mass matrix induced by the heavy states, since 
in the effective Lagrangian picture all the L violation is encoded in the single Weinberg operator, no matter what is the heavy state \cite{Antusch:2004xy}. This can be clearly seen from the one loop diagram which all have as subdiagram one of the seesaw diagram of Fig.~1.
Moreover, since both tree level and one loop diagrams, put together, form a closed loop, the L violation from the heavy state must be accompanied by L violation from the decaying heavy state, which in order to close the loop can come only through the combination of Yukawa couplings producing the neutrino masses. This implies that the asymmetry must be proportional to the neutrino mass matrix induced by the decaying state too.\footnote{Note nevertheless that obviously the decaying state, being on-shell, cannot be integrated-out, which means that the numerical prefactor in front of both neutrino mass matrices in the CP asymmetry has no reason to be the same in the 3 cases. Both the kinetic and mass terms of the decaying state contribute and $SU(2)_L$ index contractions can be different.} Therefore, for the 3 possible types of decaying particles one gets a CP asymmetry quadratic in neutrino mass matrix contributions and inversely proportional to $v^4\Gamma_D$, where $\Gamma_D$ is the total decay width of the decaying state. This means that an extra $M_D^3$ factor must come, with $M_D$ the mass of the decaying state. For instance, for the 3 seesaw cases, we get the simple form:
\begin{eqnarray}
\varepsilon_{N}&=&-\frac{3}{32\pi^2}\frac{{m}_N^3}{\Gamma_Nv^4}Im[({\cal M}^N_\nu)_{\beta\alpha}({\cal M}^H_\nu)^\dagger_{\alpha\beta}]\label{epsIhierarc}\\
\varepsilon_{\Delta}&=&-\frac{1}{16\pi^2}\frac{{m}_\Delta^3}{\Gamma_\Delta v^4}Im[({\cal M}^\Delta_\nu)_{\beta\alpha}({\cal M}^H_\nu)^\dagger_{\alpha\beta}]\quad \cite{Hambye:2003ka,Hambye:2005tk}\label{epsIIhierarc}\\
\varepsilon_{\Sigma}&=&-\frac{1}{32\pi^2}\frac{{M}_\Sigma^3}{\Gamma_\Sigma v^4}Im[({\cal M}^\Sigma_\nu)_{\beta\alpha}({\cal M}^H_\nu)^\dagger_{\alpha\beta}]\quad \cite{Hambye:2003rt}\label{epsIIIhierarc}
\end{eqnarray}
with ${\cal M}^N$, ${\cal M}^\Sigma$, ${\cal M}^\Delta$ and ${\cal M}^H$ the neutrino mass matrix induced by the decaying $N$, the decaying $\Sigma$, the decaying $\Delta_L$ and whatever heavier state in the loop diagram, respectively. For the type-I and III cases, the cubic $M_{N,\Delta,\Sigma}^3$ has an important consequence. Given the fact that $\Gamma_{N,\Sigma}$ goes parametrically as $ M^2_{N,\Sigma}{\cal M}_\nu^{N,\Sigma}/v^2$, the CP-asymmetry is linear in the mass of the lighter state (and linear in neutrino masses), which as noticed long ago \cite{Barbieri:2000ma,Buchmuller:1999zf,Hambye:2001eu,branco,Davidson:2002qv,Hambye:2003rt} is at the origin of the fact that leptogenesis in the hierarchical case requires heavy states, with upper bounds precisely determined in Ref.~\cite{Davidson:2002qv} and \cite{Hambye:2003rt} for the type I and III case respectively, see below. For the type-II case this is more complicated because the total decay width receives contributions from two different decays (to $LL$ and $HH$) but there exists also a lower bound on $m_\Delta$ \cite{Hambye:2000ui,Hambye:2005tk}, see below.

Below we will also give, in specific models, the CP asymmetries one gets when the virtual states are not much heavier than the decaying particle.
\end{itemize}

\section{Leptogenesis from the decay of a fermion triplet}

We begin by considering the decay of a fermion triplet \cite{Hambye:2003rt,Fischler:2008xm,Strumia:2008cf,AristizabalSierra:2010mv,Zhuridov:2012hb,AristizabalSierra:2012js}. This case is similar to the one of the decay of a right-handed neutrino except for one important difference: triplets can be thermalized by gauge interactions. For clarity, it is natural, as we will do here, to consider the fermion triplet possibility before the scalar triplet one because the latter option involves not only gauge interactions but also several decays, whose interplay introduces extra subtleties.

\subsection{Hierarchical case}

In this subsection, we consider the case of a fermion triplet much lighter than the other seesaw states. In this case the discussion is greatly simplified for three main reasons. First, as discussed in point (v) of the previous section, the CP-asymmetry takes in this case a universal form, no matter what are the heavy state, Eq.~(\ref{epsIIIhierarc}).
Second, the effects of the lepton number violating processes involving the heavier states can be neglected. These processes are either Boltzmann suppressed if these heavy states are on-shell in these processes, or is suppressed by power of the masses of these heavy states if they are off-shell. Third the lepton asymmetry produced by the decay of these heavier states can be generally ignored because it is washed-out by the interactions of the lightest one before this state becomes out of thermal equilibrium (although this is not always the case, especially in the flavour case, see below). 

A fermion triplet, as a right-handed neutrino, decays only through Yukawa couplings. Its neutral component is a Majorana right-handed neutrino and has consequently the same decay width as a function of these Yukawa couplings
\begin{equation}
\Gamma_{\Sigma^0}=\Gamma(\Sigma^0\rightarrow L H)+\Gamma(\Sigma^0 \rightarrow \bar{L}\bar{H})=\frac{1}{8\pi}m_\Sigma |Y_{\Sigma i}|^2\,.
\label{fermGamma}
\end{equation}
As for the charged states, the (right-handed) $\Sigma^\pm$ and the (left-handed) conjugated states of the $\Sigma^\mp$ form a Dirac spinor $\psi^\pm$ with, from SU(2)$_L$ invariance, same decay width as in Eq.~(\ref{fermGamma}).
The CP-asymmetry, given in Eq.~(\ref{epsIIIhierarc}), turns out to be 3 times smaller than for a right-handed neutrino, Eq.~(\ref{epsIhierarc}). By $SU(2)_L$ invariance, it is equal for each of the 3 triplet components, $\Sigma^0,\,\psi^+,\psi^-$.

The lepton asymmetry produced by the decay of a heavy triplet is given by the average $\Delta L$ produced per decay, $\varepsilon_\Sigma$, times the number density of triplets which have decayed, $n_\Sigma/s$, times the "efficiency factor", $\eta$,
\begin{equation}
\frac{n_L}{s}=\varepsilon_\Sigma \eta \frac{n_\Sigma}{s}\Big|_{T>>m_\Sigma}\,,
\end{equation}
with $n_\Sigma=n_{\Sigma^0}+n_{\psi^+}+n_{\psi^-}$ the total number of triplets (both particles and antiparticles).
The $\eta$ factor results from the integration of the Boltzmann equations, see below. It takes its maximum value, unity, if all decays occur out-of-equilibrium and if there is no wash-out effect from $\Delta L\neq0$ processes. It is smaller otherwise. In this way the baryon asymmetry produced is
\begin{equation}
\frac{n_B}{s}= 3 \frac{135 \zeta(3)}{4\pi^4 g_\ast}C_{L\rightarrow B}\,\varepsilon_\Sigma \eta=-0.0041 \, \varepsilon_\Sigma \,\eta \,,
\label{nBoverssigma}
\end{equation}
where $s=g_\ast(2 \pi^2/45)T^3$ is the entropy density and $C_{L\rightarrow B}=-28/79$ \cite{Harvey:1990qw} is the L to B asymmetry conversion factor one gets in the SM from the sphaleron processes. The factor of 3 in the last equation accounts for the sum of the asymmetries of the 3 triplet states. For the situation where, beside the seesaw states, there are only SM states in the thermal bath of the Universe at $T\sim M_\Sigma$ when the triplets decay, the number of relativistic degrees of freedom $g_\ast$ is equal to $106.75$.

As mentioned above, in the type-I seesaw case the CP asymmetry, Eq.~(\ref{epsIhierarc}), involves the same number of Yukawa couplings and seesaw state mass than the combination $m_N {\cal M}_\nu/v^2$.
As a result the CP-asymmetry of Eq.~(\ref{epsIhierarc}) turns out to be bounded from above as \cite{Davidson:2002qv}
\begin{equation}
\varepsilon_N\le\frac{3}{8\pi}\frac{M_{N}}{v^2}(m_{\nu_3}-m_{\nu_1})= \frac{3}{8\pi}\frac{M_{N}}{v^2}\frac{\Delta m^2_{atm}}{m_{\nu_3}+m_{\nu_1}}\,,
\label{epsIbound}
\end{equation}
which gives the bound
\begin{equation}
m_N\gtrsim 6\cdot 10^8\,\hbox{GeV}\,,
\label{MNbound}
\end{equation}
with $\Delta m^2_{atm}\simeq 2.4\cdot 10^{-3}$~eV$^2$, the atmospheric neutrino squared mass splitting \cite{GonzalezGarcia:2010er}.

Exactly the same expression holds for triplet, except for the 1/3 factor in the CP-asymmetry
\begin{equation}
\varepsilon_\Sigma\le\frac{1}{8\pi}\frac{M_{\Sigma}}{v^2}(m_{\nu_3}-m_{\nu_1})= \frac{1}{8\pi}\frac{M_{\Sigma}}{v^2}\frac{\Delta m^2_{atm}}{m_{\nu_3}+m_{\nu_1}}\,.
\label{epsIIIbound}
\end{equation}
This factor is nevertheless compensated by the factor 3 in Eq.~(\ref{nBoverssigma}). 
To calculate the lower bound it gives on $m_\Sigma$, one cannot simply assume a situation where the efficiency is unity, as one can for a right-handed neutrino decay. Gauge scatterings give a suppression of the efficiency which, to get this lower bound, must be calculated from the Boltzmann equations. 
Since different components of the fermion triplet can be involved in a same gauge scattering, it is convenient to consider the Boltzmann equation for the sum of all 3 triplet components, $Y_\Sigma \equiv n_\Sigma/s$. One gets \cite{Hambye:2003rt}
   \begin{eqnarray}
sHz \frac{dY_\Sigma}{dz} &=&
  -\bigg(\frac{Y_{\Sigma}}{Y_{\Sigma}^{\rm eq}}-1\bigg)\gamma_D
  -2 \bigg(\frac{Y_{\Sigma}^2}{Y_{\Sigma}^{2\rm eq}}-1\bigg)\gamma_A \,,
  \label{BoltzSigma} \\
sHz \frac{dY_{{B} - {L}}}{dz} &=&
-\gamma_D \varepsilon_{\Sigma} \bigg(\frac{Y_{\Sigma}}{Y_{\Sigma}^{\rm eq}}-1\bigg)  
-\frac{Y_{{B} - {L}}}{Y_{l}^{\rm eq}}\bigg(\frac{\gamma_D}{2}+
2\gamma_{\Sigma}^{\rm sub}\bigg) \,.
  \label{BoltzB-L}
\end{eqnarray}
In these equations $z\equiv m_\Sigma/T$ is the evolution variable, $H=1.66 \sqrt{g_\star} T^2/m_{Pl}$ is the Hubble constant and the suffix $eq$ denotes the equilibrium value. $Y_l^{eq}$ stands for the equilibrium number of a two degrees of freedom fermion ($Y_l^{eq}=Y_\gamma$). The $\gamma_i$ are the reaction densities for the various processes (i.e.~the number of reactions  occuring per unit volume per unit time).
$\gamma_D= n_\Sigma^{eq}\Gamma_\Sigma K_1(z)/K_2(z)$ is the decay/inverse decay reaction rate.
$\gamma_\Sigma^{sub}$ comes from the $\Delta L=2$ scattering processes, $LL\leftrightarrow H^* H^*$ and $LH\leftrightarrow \bar{L} H^*$, see Ref.~\cite{Hambye:2003rt}.
$\gamma_A$ is the gauge scattering reaction density from the $\Sigma \bar{\Sigma}' \leftrightarrow, GG',\,f \bar{f},\, H\bar{H}$ processes, with $G^{(')}$ all possible gauge bosons. 
Summing over the 12 SM fermions we get the following reduced cross section \cite{Hambye:2003rt,Cirelli:2007xd,Strumia:2008cf} 
\begin{equation}
\hat\sigma_A= 
\frac{6g^4}{72 \pi}\Big[\frac{45}{2} \beta -\frac{27}{2}\beta^3-(9(\beta^2-2)+18(\beta^2-1)^2)\ln \frac{1+\beta}
{1-\beta}\Big]\,,
\end{equation}
with $\beta= \sqrt{1-4/x}$, $x=s/m^2_\Sigma$,
\begin{equation}
\gamma(a\,b\leftrightarrow 1\,2) =  \, \frac{T}{64~\pi^4} \int_{s_{min}}^{\infty} ds ~\sqrt{s}~\hat{\sigma}(s)~K_1\left(\frac{\sqrt{s}}{T} \right)\,,
\label{ScatRates}
\end{equation}
and $\hat{\sigma}\equiv  2 s^{-1} \lambda^2[s,m^2_a,m^2_b] \sigma(s)$ with $\lambda[a,b,c]=\sqrt{(a-b-c)^2- 4 b c}$.
Since the gauge scatterings do not violate L, they appear only in the $Y_\Sigma$ Boltzmann equation, Eq.~(\ref{BoltzSigma}).

In absence of the $\gamma_A$ term the efficiency is exactly the same as for the right-handed neutrinos, except that all interaction rate terms are multiplied by 3, in Eq.~(\ref{BoltzB-L}). This holds also for the inverse decay term since a lepton has 3 times more probability to encounter a Higgs particle to produce a heavy triplet (i.e.~it can inverse decays in three ways instead of one). For a given value of 
\begin{equation}
\tilde{m}=\frac{1}{2} \sum_j |Y_{\Sigma_{1j}}|^2 \frac{v^2}{m_\Sigma}=8 \pi \Gamma_\Sigma \frac{v^2}{m_\Sigma}\,,
\label{mtildedef}
\end{equation}
the efficiency suppression due to inverse decay is therefore the same as the one obtained for a right-handed neutrino with a $\tilde{m}$ value 3 times larger.

\begin{figure}[!t]
\centering
\includegraphics[height=7.5cm]{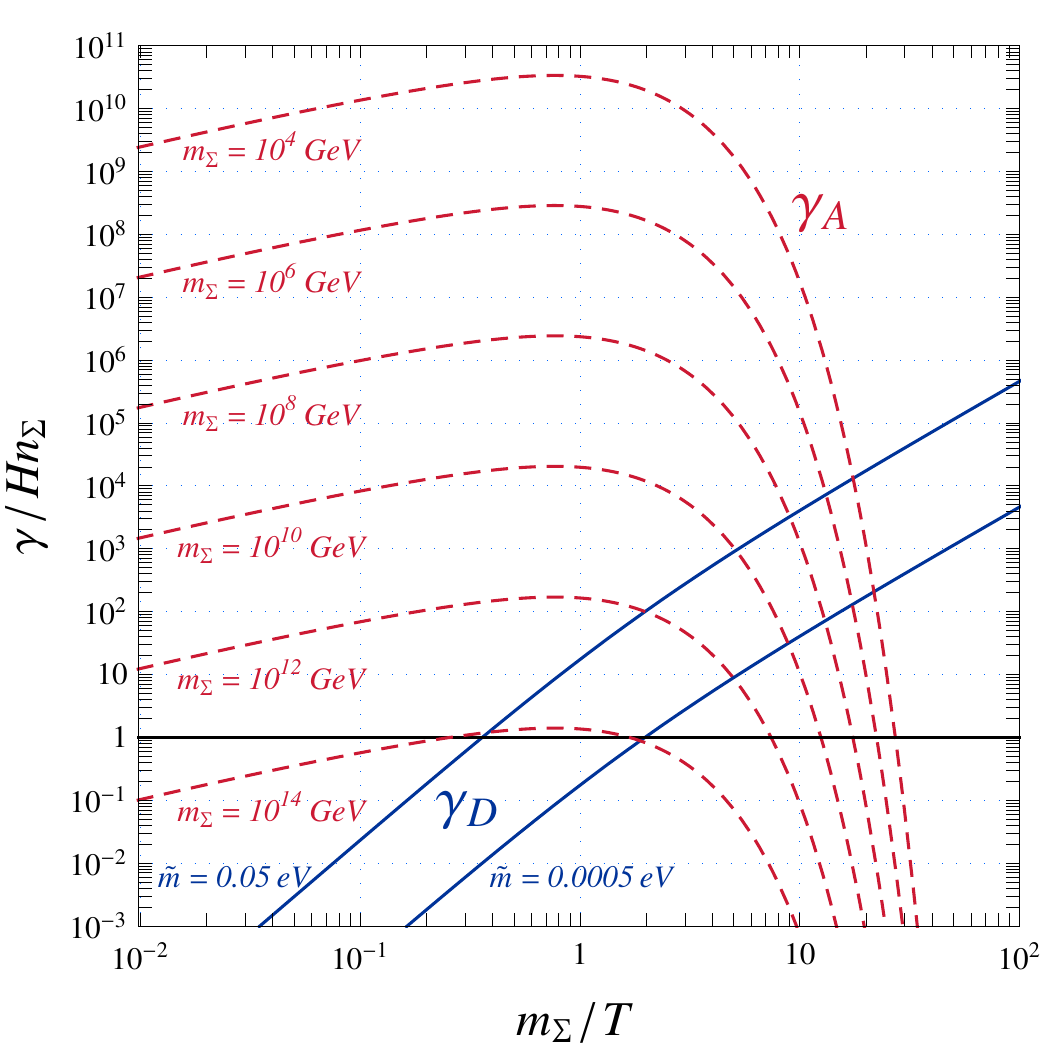}\includegraphics[height=7.5cm]{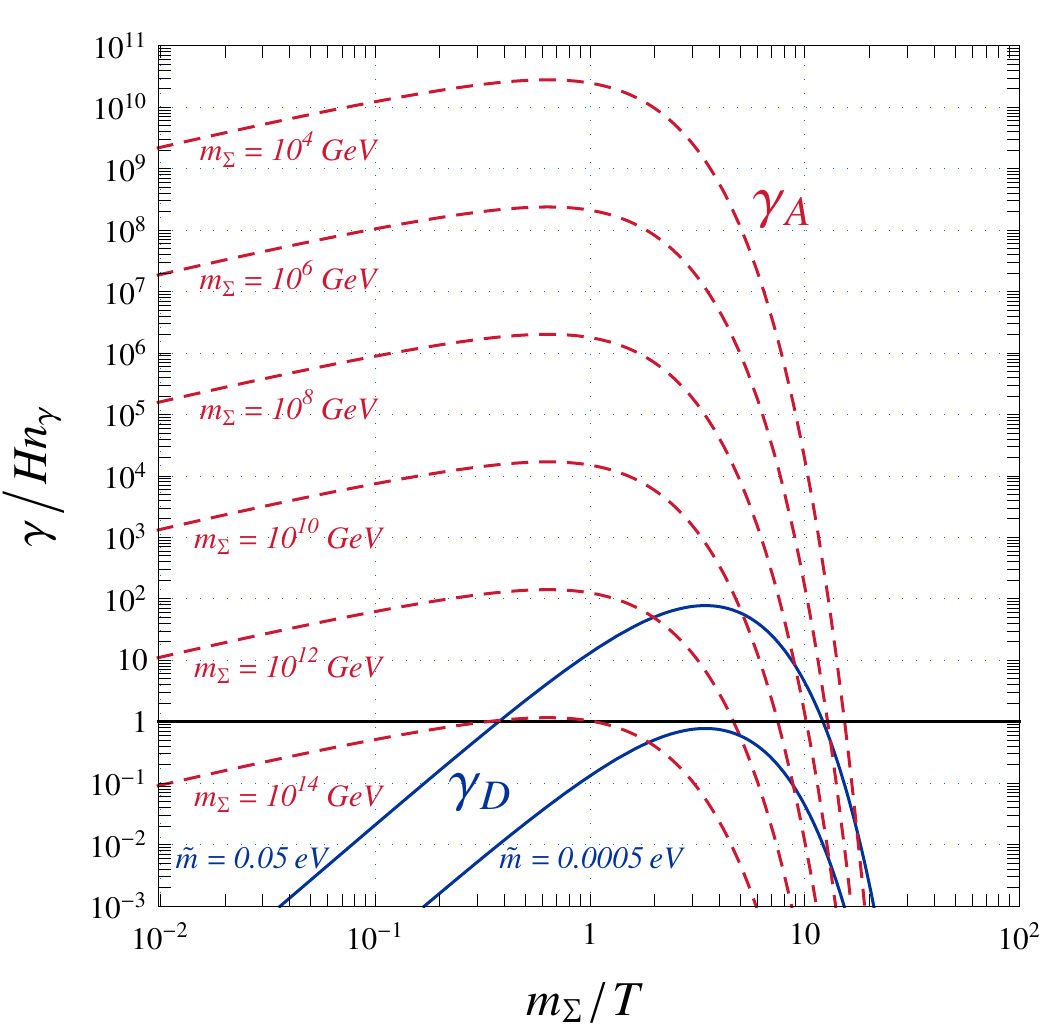}\\
\caption{$\gamma/Hn_\Sigma^{eq}$ and $\gamma/Hn_\gamma$ thermalization rates, as relevant for Eqs.~(\ref{BoltzSigma}) and (\ref{BoltzB-L}) respectively. $\gamma_D/Hn_{\Sigma,\gamma}^{eq}$ depends only on $\tilde{m}$, whereas  $\gamma_A/Hn_{\Sigma,\gamma}^{eq}$ depends only on $m_\Sigma$.}
\label{figrates}
\end{figure}

Let us now discuss in details what happen's when one adds the $\gamma_A$ gauge term.
The gauge processes do not bring any new unknown parameter.
They are fully known as a function of $m_\Sigma$. On the contrary they render leptogenesis more predictive since they can thermalize the decaying triplet, giving a baryon asymmetry independent of the initial condition on the triplet population. At large $T>>m_\Sigma$, gauge interactions are not in thermal equilibrium because dimensionally, from the Planck scale dependence of the Hubble constant, $\gamma_A/n_\Sigma^{eq}H$ necessarily goes like $m_{Pl}/T$. Similarly at $T<<m_\Sigma$ one has \cite{Strumia:2008cf} 
\begin{eqnarray}
\hat{\sigma}_A&=&c_s \beta +c_p\beta^3 +{\cal O}(\beta^5)\,,   \label{hatsigmasp}\\
\gamma_A&=&\frac{m_\Sigma T^3}{32 \pi^3} e^{-2m_\Sigma/T} \Big[ c_s+\frac{3T}{2m_\Sigma}(c_p+\frac{c_s}{2})+{\cal O}(T/m_\Sigma)^2\Big]\,.  \label{gammasigmasp}
\end{eqnarray}
where $c_s=111g^4/8\pi$ and $c_p={51g^4}/{8 \pi}$ refer to the $s$ and $p$ wave components.
The gauge reaction rate is doubly Boltzmann suppressed, since it involves 2 external heavy particles,  which gives a thermalization rate $\gamma_A/n_\Sigma^{eq}H\propto e^{-m_\Sigma/T} m_{Pl}/\sqrt{m_\Sigma T}$ which has a single Boltzmann suppression. As a result  $\gamma_A/n_\Sigma^{eq}H$ reaches a maximum value at $T\sim m_\Sigma$, which turns out to be  
\begin{equation}
\frac{\gamma_A}{n_\Sigma^{eq}H}\sim 2 \cdot \frac{10^{14}\,\hbox{GeV}}{m_\Sigma}\,.
\end{equation}
Therefore as soon as $m_\Sigma$ is below $\sim 10^{14}$~GeV the triplet thermalizes and the asymmetry produced is independent of the initial condition on the triplet population.  And the smaller is $m_\Sigma$ the more the triplet is thermalized. Actually one could believe that, as a consequence, leptogenesis becomes quickly hopeless below this value. As was shown in Refs.~\cite{Hambye:2000ui,Hambye:2003rt,Hambye:2005tk} this is nevertheless not the case at all, because, for $T<m_\Sigma$, the gauge interaction rate, which is doubly Boltzmann suppressed, drops more quickly than the decay/inverse decay $\gamma_D$ rate which is simply Boltzmann suppressed.
Therefore gauge processes decouple faster than the $\gamma_D$ term, still leaving the possibility to create a substantial asymmetry from the decay once the gauge scatterings become slower than the decay/inverse decay processes (or slightly before, as we will see below). All this can be seen in Fig.~\ref{figrates} which shows the reaction rates for various values of $m_\Sigma$ and $\tilde{m}$. Note that the $\gamma/n_\Sigma^{eq}H$ rates are the relevant ones for Eq.~(\ref{BoltzSigma}) whereas the $\gamma/n_\gamma^{eq}H$ are the relevant ones for Eq.~(\ref{BoltzB-L}). If they are smaller than unity the corresponding interactions do not thermalize in the corresponding Boltzmann equation and can be neglected. Fig.~\ref{figasymevol} gives the corresponding asymmetry we get, as a function of $z$, once we integrate the Boltzmann equations. To better understand these results let's proceed step by step in the following way:
\begin{figure}[!t]
\centering
\includegraphics[height=7.5cm]{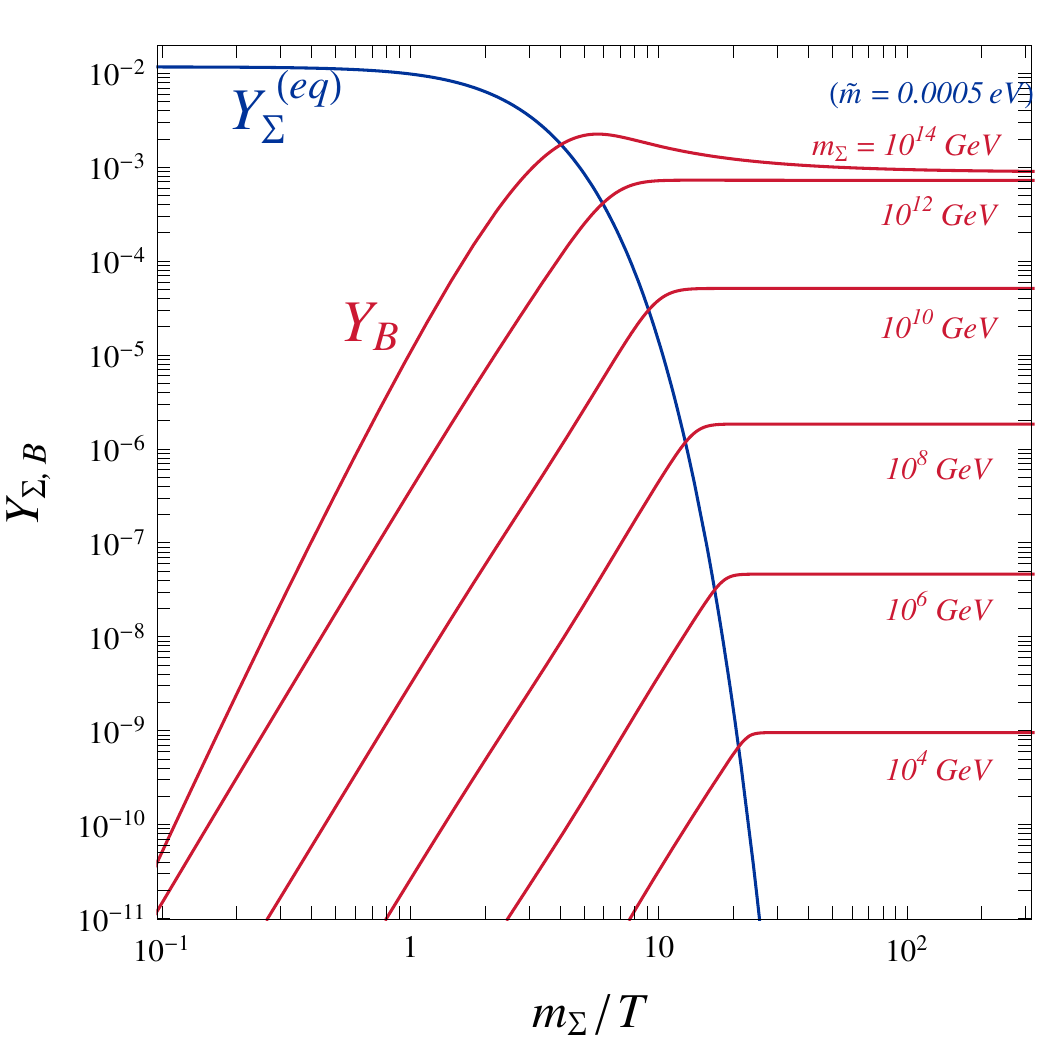}\includegraphics[height=7.5cm]{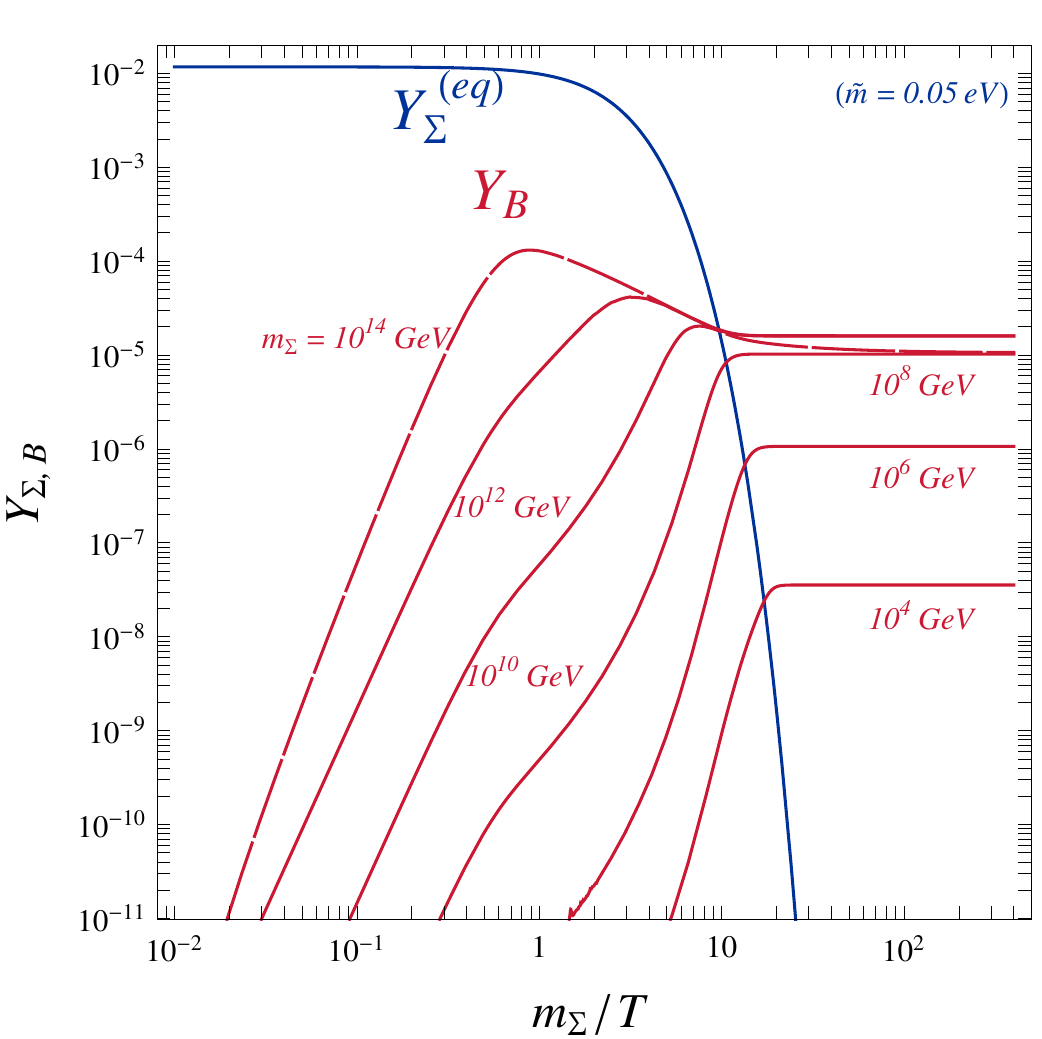}\\
\caption{$Y_\Sigma^{eq}$ and $Y_B$ as a function of $z=m_\Sigma/T$, for various values of $m_\Sigma$ and $\tilde{m}=0.0005$~eV (left panel) and $\tilde{m}=0.05$~eV (right panel). The CP-asymmetry has been taken equal to unity. For the first panel one lies always in the gauge regime, except for $m_\Sigma=10^{14}$~GeV where the suppression effect comes from the $\Delta L=2$ scattering processes. For $\tilde{m}=0.05$~eV instead, one lies in the gauge regime below $\sim10^{7}$~GeV, see Fig.~\ref{figregimes}. Above this mass value the Yukawa regime leads to an asymmetry more and more independent of $m_\Sigma$, except above $\sim10^{14}$~GeV (due to the $\Delta L=2$ scattering processes here too).}  
\label{figasymevol}
\end{figure}

\begin{itemize}
\item[(a)] To begin with, let us neglect all terms in the Boltzmann equations except the first term of both Boltzmann equations, i.e.~no gauge scattering term, no inverse decay washout term and no $\Delta L=2$ scattering terms.
In this case, no matter what the value of $\gamma_D$ is, by putting the first Boltzmann equation into the second one, one gets obviously $Y_L=\int \varepsilon_\Sigma (dY_\Sigma/dz) dz=\varepsilon_\Sigma Y^{eq}_\Sigma|_{T>>m_\Sigma}$, that is to say an efficiciency equal to unity.
\item[(b)] Adding now the gauge term, clearly as long as  $\gamma_A/n_\Sigma^{eq}H>1$, $Y_\Sigma$ tracks $Y_\Sigma^{eq}$. Moreover as long as $4\gamma_A> \gamma_D$, the gauge term dominates the thermalization of the triplet (i.e.~the triplets scatter before they decay), see Eq.~(\ref{BoltzSigma}). In this case putting Eq.~(\ref{BoltzSigma}) in Eq.~(\ref{BoltzB-L}) one simply gets
\begin{equation}
\frac{dY_L}{dz}=\varepsilon_\Sigma \frac{dY_\Sigma^{eq}}{dz} \frac{\gamma_D}{4\gamma_A}\,.
\end{equation}
This equation remains correct until a value of $z=z_A$ where either $\gamma_A/n_\Sigma^{eq}H$ gets below one (in case the gauge scattering are no longer in thermal equilibrium) or $4\gamma_A$ gets below $\gamma_D$. As a result the lepton asymmetry contains 2 pieces: a lepton asymmetry produced when $z\lesssim z_A$, which is suppressed by the $\gamma_D/4\gamma_A$ factor, and an asymmetry produced afterwards, when $z\gtrsim z_A$. The latter is not suppressed by any wash-out effect, i.e.~all triplets left at $z=z_A$ produce leptons without any suppression, but it is obviously suppressed by the fact that the number of triplets left at $z=z_A$ is Boltzmann suppressed. Summing these 2 contributions one gets
\begin{equation}
Y_L\simeq \varepsilon_\Sigma \int_{z_{in}}^{z_A} \frac{dY^{eq}_\Sigma}{dz}\frac{\gamma_D}{4\gamma_A} +\varepsilon_\Sigma Y^{eq}_\Sigma(z_A)\,.
\label{YLsigma1}
\end{equation}
It is important to stress that unless $\tilde{m}$ is quite small, $\gamma_D/(n_{\Sigma}^{eq}H)$ (which is not a Boltzmann suppressed quantity!) is larger than one when $\gamma_A$ gets Boltzmann suppressed, at $z\sim$~a~few. Thus, the value of $z_A$ is determined by the condition $4 \gamma_A=\gamma_D$ rather than by the condition $\gamma_A/(n_{\Sigma}^{eq}H)=1$.
This means that the gauge scatterings become irrelevant already before they get out-of-equilibrium, i.e.~for $z>z_A$ the triplets decay before they scatter even if their scattering rate is still fast. That is precisely the reason why the suppression effect of the efficiency from the gauge scatterings is not as dramatic as we could have expected at first sight.
Note also that in the integrand of Eq.~(\ref{YLsigma1}) the various Boltzmann suppressions of the various terms cancel each other, leaving an integrand which scales as $1/T^3$. As a result we get
\begin{equation}
Y_L\simeq \varepsilon_\Sigma Y^{eq}(z_A)\int_{z_{in}}^{z_A} \frac{z^3}{z_A^3} dz+\varepsilon_\Sigma Y_\Sigma^{eq}(z_A)\simeq \varepsilon_\Sigma Y_\Sigma^{eq}(z_A) (z_A/4+1)\,.
\label{YLsigma2}
\end{equation}
Since $z_A=$~a few, that is to say lies between $\sim1$ for $m_\Sigma\simeq10^{14}$~GeV and $\sim 25$ for $m_\Sigma\simeq 1$~TeV, this means that a good part of the asymmetry is produced just before $z=z_A$ and the rest just after. This can be seen in Fig.~\ref{figasymevol}.a  which, with $\varepsilon_\Sigma=1$, gives $Y_L$ of order $Y_\Sigma^{eq}(z=z_A)$ (and similarly in Fig.~\ref{figasymevol}.b for low enough $m_\Sigma$). In all cases note that the smaller $\tilde{m}$ is, the smaller $\gamma_D/\gamma_A$ is, the larger $z_A$ is, the more $Y_\Sigma^{eq}(z=z_A)$ is Boltzmann suppressed, and hence the more $Y_L$ is Boltzmann suppressed.

\begin{figure}[!t]
\centering
\includegraphics[height=5.9cm]{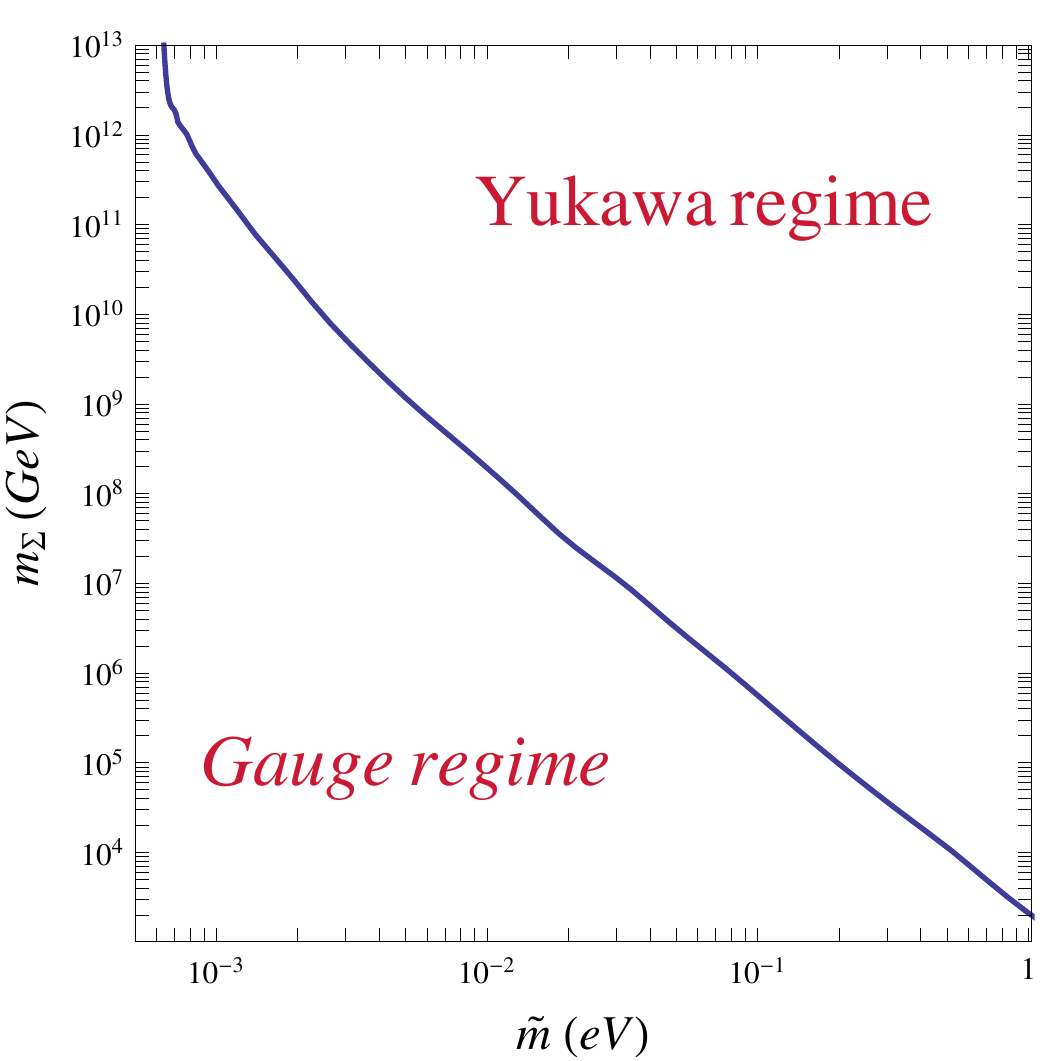}\\
\caption{Gauge versus Yukawa regime. In the gauge regime the inverse decay washout term can be neglected and the asymmetry is given by Eq.~(\ref{YLsigma2}). In the Yukawa regime instead, the gauge annihilation terms can be neglected and one get an efficiency which has the same behavior than for a decaying right-handed neutrino. Only close to the transition line, which gives the maximum asymmetry for fixed $m_\Sigma$, can both terms be relevant.}
\label{figregimes}
\end{figure}

\item[(c)] Now let us add the $\Delta L=2$ scattering term, $\gamma_\Sigma^{sub}$. As in the type-I framework these terms are proportional to the seesaw mass scale, $m_\Sigma$, and, since they involve the neutrino mass Feynman diagram, to the neutrino mass scale. For neutrino masses in agreement with the data (i.e.~below the $\sim$eV scale) their effect is small for $m_\Sigma$ below a scale which depends on $\tilde{m}$.
For a hierarchical or inverse hierarchical neutrino mass spectrum, with $\tilde{m}\lesssim 10^{-1}$~eV there is no effect below $\sim10^{14}$~GeV whereas, for $\tilde{m}\sim 1$~eV, there is a sizeable effect down to $10^{12}$~GeV. For a quasi-degenerate spectrum of light neutrino the mass where they become to have a sizeable effect scales as the inverse of the neutrino mass scale.

\item[(d)] Let us now add the $\propto Y_{B-L} \gamma_D$ inverse decay washout term in Eq.~(\ref{BoltzB-L}). What is its effect? Well, the relevant quantity for the inverse decay washout is the $\gamma_D/(n_{\gamma}H)$ thermalization rate (which unlike $\gamma_D/(n_{\Sigma}^{eq}H)$ is Boltzmann suppressed at $z>1$!).
Therefore, 
 if at $z=z_A$, the inequality $\gamma_D/(n_{\gamma} H)<1$ holds, clearly Eq.~(\ref{YLsigma2}) remains valid. This is true clearly if $\tilde{m}\lesssim  3\cdot 10^{-3}$~eV since in this case $\gamma_D/(n_{\gamma} H)$ never reaches unity. More generally this is also true for larger values of $\tilde{m}$, since at $z=z_A>1$,  $\gamma_D/(n_{\gamma} H)$ is already Boltzmann suppressed.
The value of $\tilde{m}\equiv \tilde{m}_{A}$ which gives 
\begin{equation}
\frac{\gamma_D}{n_{\gamma} H}\Big|_{z=z_A}=1
\end{equation}
is given in Fig.~\ref{figregimes}.
It delimitates the lower $\tilde{m}$ "gauge regime" region where inverse decays can be neglected, so that Eq.~(\ref{YLsigma2}) holds, from the larger $\tilde{m}$ "Yukawa regime" region where inverse decays are important. 
These two regions present a totally different behaviour of the efficiency as a function of $\tilde{m}$, see Fig.~\ref{mtildeetaIIIbis} (a similar plot can be found in Refs.~\cite{Fischler:2008xm,AristizabalSierra:2010mv}). In the gauge regime, as explained above, the efficiency increases with $\tilde{m}$. In the Yukawa regime instead the decay/ inverse decay processes dominate and the efficiency decreases with $\tilde{m}$ (similarly to what happen's in the type-I scenario). Consequently a maximum efficiency is obtained at the junction between both regime, for $\tilde{m}\simeq \tilde{m}_A$.
Note that at this maximum the gauge scattering processes have still nevertheless a suppression effect, but it is already moderate, of order $\simeq 3$-$5$.
Consequently, as Fig.~\ref{mtildeetaIIIbis} shows, at this maximum point the efficiency still depends on $m_\Sigma$. The gauge scattering effect, and hence the $m_\Sigma$ dependence, becomes to be totally negligible for a value of $\tilde{m}$ equal to about 10 times $\tilde{m}_A$, see Fig.~\ref{mtildeetaIIIbis}. At this point even if the gauge scatterings have still an effect at early times, they don't affect the final asymmetry produced. 
This is similar to the known feature of the type-I leptogenesis scenario, that in the strong washout regime (i.e.~for $\tilde{m}>>10^{-3}$~eV), the final lepton asymmetry produced doesn't depend on the reheating temperature as soon as this temperature is somewhat larger than the temperature where the inverse decays decouple. For $\tilde{m}>\tilde{m}_{A}$ one is therefore in a regime totally dominated by the Yukawa interaction, as with right-handed neutrinos.\footnote{This separation between Yukawa and gauge regimes has been already quantified analytically in Ref.~\cite{AristizabalSierra:2010mv}, taking the different prescription for defining $\tilde{m}_A$ that $\gamma_D/(n_{\gamma} H)=1$ when $\gamma_A/(n^{eq}_{\Sigma} H)=1$. This results in values of $\tilde{m}_{A}$ which, for low values of $m_\Sigma$, can be orders of magnitudes larger. For example for $m_\Sigma=10^6$~GeV our prescription gives $\tilde{m}_A\simeq 0.1$~eV, to be compared with the value $\tilde{m}_A=25$~eV one gets adopting the prescription of Ref.~\cite{AristizabalSierra:2010mv}. Fig.~\ref{figasymevol}.b shows in this case that the lepton asymmetry is produced when $\gamma_A\simeq \gamma_D$, which occurs at $z\simeq 13$, when $\gamma_A\simeq \gamma_D$, see Fig.~\ref{figrates}, rather than when $\gamma_A/(n^{eq}_{\gamma} H)=1$ which occurs at $z=20$ (where all quantities are much more Boltzmann suppressed). This doesn't affect nevertheless the numerical values of $Y_L$ obtained in Ref.~\cite{AristizabalSierra:2010mv} with whom we agree. The prescription of Ref.~\cite{AristizabalSierra:2010mv} holds for a situation where the effects of gauge scatterings are already very suppressed, see Fig.~\ref{mtildeetaIIIbis}.}

\begin{figure}[!t]
\centering
\includegraphics[height=5.9cm]{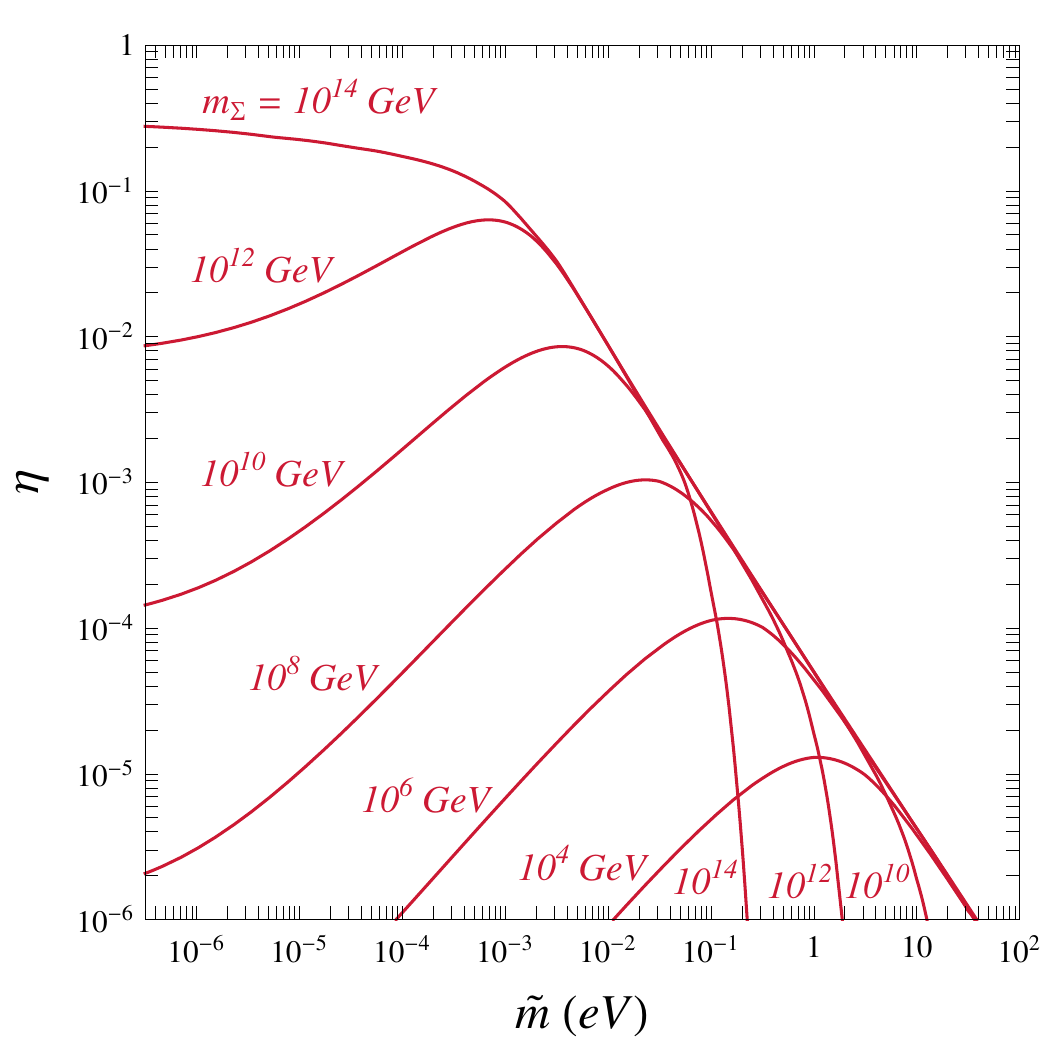}\\
\caption{Efficiency as a function of $\tilde{m}$ for various values of $m_\Sigma$.}
\label{mtildeetaIIIbis}
\end{figure}

\end{itemize}

\begin{figure}[!t]
\centering
\includegraphics[height=6.5cm]{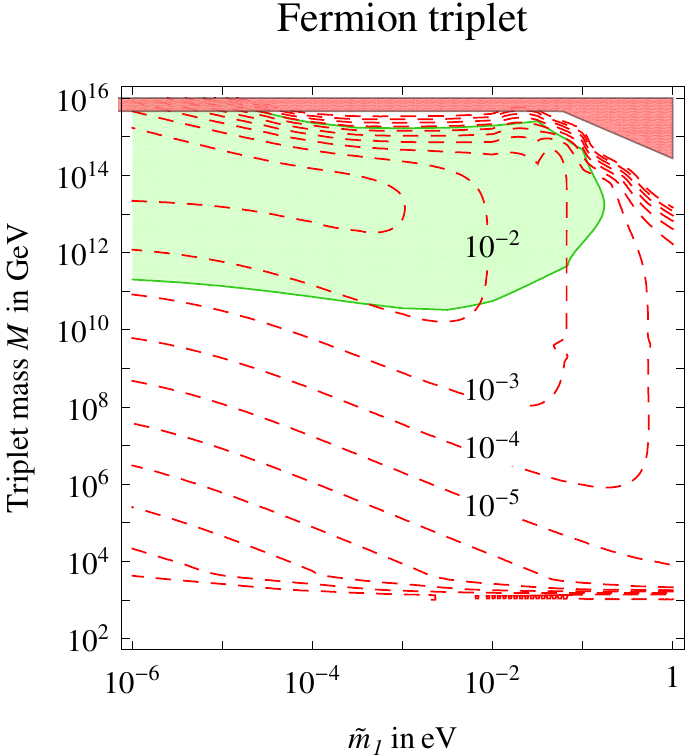}\\
\caption{Efficiency for a decaying fermion triplet, from Ref.~\cite{Strumia:2008cf}. Note that this figure differs from a similar one obtained in Ref.~\cite{Hambye:2003rt} by the fact that Sommerfeld enhancement effects for non-relativistic triplets in gauge scatterings have been taken into account. These effects modify the final baryon asymmetry by less than 20$\%$.} 
\label{figeffic}
\end{figure}

All the properties above can also be seen in Fig.~\ref{figeffic} which shows a contour plot of the efficiency one obtains, as a a function of $\tilde{m}$ and $m_\Sigma$. Look in particular at Fig.~\ref{figregimes} to understand the change of efficiency behavior in Figs.~6 and~\ref{figeffic}. These figures also show that the efficiency quickly drops for $m_\Sigma$ below a few TeV or above $\sim10^{12-14}$~GeV (depending on the value of $\tilde{m}$). In the first case this is due to the sphalerons  which get out-of-equilibrium at the electroweak scale. The value of the decoupling temperature one gets for $m_h=125$~GeV is $T_c=(139.5\pm 2.5)$~GeV, see Fig.~3 of Ref.~\cite{Burnier:2005hp}. It is a good approximation to simply use a sharp step function for this decoupling. In the second case this is due to the fact that above $\sim10^{12-14}$ GeV the Yukawa induced $\Delta L=2$ scattering processes (which are proportional to $m_\Sigma$) cease to have a negligible wash-out effect.

Fig.~\ref{figeffic} also shows the region of parameter space which, from the CP asymmetry one gets in the hierarchical case, Eq.~(\ref{epsIIIhierarc}), gives a baryon asymmetry large enough to account for the observed one. It gives the lower bound \cite{Hambye:2003rt,Fischler:2008xm,Strumia:2008cf}
\begin{equation}
m_\Sigma\gtrsim 3\cdot 10^{10}\,\hbox{GeV}\,,
\label{msigmabound}
\end{equation}
which, due to the gauge scatterings, is about 2 orders of magnitude larger than for the decay of a right-handed neutrino, Eq.~(\ref{MNbound}). 
Taking into account the fact that the CP asymmetry is linear in $m_\Sigma$ one can understand easily this bound from the discussion above. This bound is obtained for $\tilde{m}\simeq \tilde{m}_A$ which for this value of the triplet mass has the value $2.2\cdot 10^{-3}$ with $z_A=7.2$.
Thus, in this case the gauge scatterings suppress the asymmetry produced by a factor $\sim Y_\Sigma^{eq}(z=z_A)/Y_\Sigma^{eq}(z<<1)=1.3\cdot 10^{-2}$, that is to say by 2 orders of magnitude. That must be compensated by a 2 order of magnitude $m_\Sigma$ increase in the CP asymmetry, Eq.~(\ref{epsIIIbound}).
As for an upper bound on $m_\Sigma$, it is of order $10^{16}$~GeV and is not far from the one we would get from requiring perturbative Yukawa coupling (if the decaying triplet gives a neutrino mass of order the atmospheric value 0.05~eV).

Finally note that in the supersymmetric context, where the fermion triplet comes together with its scalar superpartner, all the discussion above remains unchanged, up to factor of order unity for the CP-asymmetries and efficiency factor. Eq.~(\ref{msigmabound}) holds in this case up to a factor of order unity (which to our knowledge has not been determined exactly in the literature). This bound implies a leptogenesis scale above the upper bound that in the supersymmetric context, the gravitino implies on the reheating temperature, $T_{reh}\lesssim 10^{6-10}$~GeV \cite{Khlopov:1984pf,Kawasaki:2008qe,Giudice:2008gu}. This basically means that triplets with a mass large enough to insure successful leptoenesis would never have been created in the thermal bath of the Universe.
The bound of Eq.~(\ref{msigmabound}) can nevertheless be relaxed by orders of magnitude if there is more than one triplet fermion and if they have a similar mass, see next subsection. One could also think about possible additional effects from supersymmetry breaking soft terms involving the triplets. For a recent extensive review on type-I seesaw soft leptogenesis \cite{Boubekeur:2002jn}-\cite{softlepto7} see Ref.~\cite{Fong:2011yx}. To our knowledge, for the type-III seesaw framework, this possibility has also not been explored in details in the literature.

\subsection{Resonant case: multiple fermion triplets with quasi degenerate spectrum }

The bound of Eqs.~(\ref{epsIIIbound}) and (\ref{msigmabound}) are strictly valid only in the limit where the mass of the heaviest seesaw states goes to infinity (with neutrino masses fixed). For a less hierarchical case, for example if the seesaw state masses differ by a factor 10, these bounds can be relaxed for particular structures of the Yukawa and seesaw state mass matrix, see Ref.~\cite{Hambye:2003rt}. In this reference this has been discussed for the right-handed neutrino case. This discussion also applies to the fermion triplet case.
For a quasi-degenerate spectrum instead the all discussion changes because in this case the self-energy diagram involving several triplets,  Fig.~\ref{fig2}, becomes resonant, through the propagator of the virtual triplet (for references on resonant leptogenesis in the type-I model, see e.g.~Refs.~\cite{resonant}-\cite{resonant13}). With several fermion triplets the CP-asymmetry of, say, $\Sigma_1$, takes the form \cite{Hambye:2003rt}
\begin{equation}
\varepsilon_{\Sigma_1}=-\sum_j \frac{3}{2}\frac{m_{\Sigma_1}}{m_{\Sigma_j}}\frac{\Gamma_{\Sigma_j}}{m_{\Sigma_j}}I_j\frac{V_j-2S_j}{3}\,,
\label{epsIIIresonant}
\end{equation}
with 
\begin{equation}
S_j=\frac{m^2_{\Sigma_j} \Delta m^2_{1j}}{(\Delta m^2_{1j})^2+m_{\Sigma_1}^2 \Gamma_{\Sigma_j}^2},\quad V_j=2\frac{m_{\Sigma_j}^2}{m_{\Sigma_1}^2}\Big[(1+\frac{m_{\Sigma_j}^2}{m_{\Sigma_1}^2}) \log(1+\frac{m_{\Sigma_1}^2}{m_{\Sigma_j}^2})-1\Big]\,,
\end{equation}
and
\begin{equation}
I_j=\frac{\hbox{Im}[(Y_\Sigma Y_\Sigma^\dagger)_{1j}^2}{|Y_\Sigma Y_\Sigma^\dagger|_{11}|Y_\Sigma Y_\Sigma^\dagger|_{jj}}, \quad\quad  \Delta m^2_{ij}=m^2_{\Sigma_j}-m^2_{\Sigma_i}\,.
\end{equation}
In the hierarchical limit both the self energy and vertex diagram contributions take the value unity, $S_j=V_j=1$.
In the extreme opposite case where the mass splitting would be equal to the decay width of the intermediate triplet, $m_{\Sigma_j}-m_{\Sigma_1}\simeq \Gamma_{\Sigma_j}$, the resonance is maximum, $S_j\simeq m_{\Sigma_j}/\Gamma_{\Sigma_j}$, leading to a CP asymmetry which can be as large as 1/2 \cite{Hambye:2003rt}. Together with the CP-asymmetry of the second quasi-degenerate triplet, one obtains a maximum CP-asymmetry equal to unity, which can lead to successful leptogenesis provided the following lower bound is satisfied  \cite{Strumia:2008cf}
\begin{equation}
m_\Sigma>1.6\,\hbox{TeV}\,.
\label{msigmaboundtev}
\end{equation}
This bound results from the interplay of the gauge scatterings, which for $m_\Sigma\sim 1$~TeV allows a production of the asymmetry only for $z>15$--$20$, and of the sphaleron decoupling scale which forbids any sizable production below it. 

Finally note that for the type-III framework, as for the type-I framework, there exist upper bounds on neutrino masses from the constraint of successful leptogenesis, but none of them appears to be much relevant, when compared to the other bounds which exist already on the neutrino mass scale, for example the direct experimental bound from the Mainz experiment, $m_{\nu}^{max}<2.2$~eV \cite{Weinheimer:1999tn}.
For instance, it is true that, as in the type-I case \cite{Buchmuller:2002jk,Buchmuller:2003gz,Giudice:2003jh,Hambye:2003rt}, an "infinite" hierarchy between the triplet seesaw states do lead to an upper bound \cite{Hambye:2003rt}
\begin{equation}
m_\nu<0.12\,\hbox{eV}\,.
\end{equation}
This bound results from 2 effects which suppress the asymmetry when the neutrino mass scale increases. First the upper bound on the CP-asymmetry decreases with this mass scale, Eq.~(\ref{epsIIIbound}). Second in the type-III model, as in the type-I model, the $\tilde{m}$ parameter is bounded from below by the value of the lightest neutrino mass
\begin{equation}
\tilde{m}\geq m_\nu^{min}\,.
\label{mtildemintypeI}
\end{equation}
This inequality results from the fact that $\tilde{m}$, Eq.~(\ref{mtildedef}), and the neutrino mass matrix seesaw formula, Table 1, differ only by a flavour index summation and absolute values on the Yukawa couplings. However given the structure of the seesaw formula, ${\cal M}_{\alpha\beta}\propto \sum_{N_i} Y_{N_{i\alpha}}Y_{N_{i\beta}}/m_{N_i}$, it doesn't appear at all easy to get a quasi-degenerate spectrum of light neutrinos, which applies for this bound, from a hierarchical spectrum of heavy triplets. That would require a precise cancellation between the hierarchy of heavy state masses and the hierarchy of Yukawa couplings. To get a quasi-degenerate spectrum of light neutrinos it is much easier to assume that the heavy states would be themselves quasi-degenerate. In this case, as discussed above, a resonance effect occurs, and there is no more upper bound on neutrino masses for successful leptogenesis \cite{Hambye:2003rt} (and this even without talking about flavour effects which further relax these bounds, even for hierarchical states \cite{Abada:2006ea}). 

\subsection{Effect of flavour}

Above we have made the assumption that the 3 Boltzmann equations for  each $L_{e,\mu,\tau}$ flavour lepton number can be reduced to a single Boltzmann equation, for the sum of them, i.e.~for the total lepton number. As well-known this is fully justified for a decaying right-handed neutrino with a mass above $10^{12-13}\,\hbox{GeV}$. However, below this scale this is not necessarily justified because the $\tau$ Standard Model Yukawa coupling interaction is in thermal equilibrium. As a result one gets 2 Boltzmann equations, for $Y_{L_\tau}$ and $Y_{L_\mu}+Y_{L_e}$, which summed do not reduce to the single total lepton number one. Similarly for $T$ below $\sim 10^9$~GeV the $\mu$ Yukawa coupling is also in thermal equilibrium and one gets three independent $Y_{L_{e,\mu\,\tau}}$ Boltzmann equations.
For a discussion of the flavour effects in the type-I seesaw model see e.g.~Refs.~\cite{Abada:2006ea,Nardi:2006fx,aba06a,Blanchet:2006ch,Davidson:2008bu}. For earlier references on the subject see \cite{Barbieri:2000ma,Pilaftsis:2004xx}. Flavour effects bring in general a moderate correction to the lepton flavour "aligned" case, but not always.
Two clear examples are in particular: 
\begin{itemize}
\item[-] a) \underline{Flavour hierarchy}: a substantially less suppressed efficiency can be obtained if the lightest triplet has in a same lepton flavour channel a large CP asymmetry and a small decay width (i.e.~typically with $\varepsilon_{\Sigma}^{l_j}>>\varepsilon_{\Sigma}^{l_i}$ and $\Gamma(\Sigma\rightarrow L_i H)>>\Gamma(\Sigma\rightarrow {L}_j{H})$). Note that the neutrino mass matrix CP-violating phases which have no effect on leptogenesis in the unflavoured case can now contribute \cite{Blanchet:2006be,Pascoli:2006ie,Anisimov:2007mw}.
\item[-] b) \underline{$N_2$ leptogenesis}: if the second lightest seesaw state produces an asymmetry, this asymmetry can eventually not be erased by the lightest one. This doesn't allow to reduce the leptogenesis scale but allows to have a "spectator" triplet well below the bounds above. 
This situation can already work in the unflavoured case, obviously if all couplings of the lightest states are suppressed. In the flavour case this can work also \cite{Vives:2005ra,Blanchet:2008pw,Engelhard:2006yg} if some of the couplings of the lightest seesaw state are large, as long as the coupling to the flavour in which the asymmetry has been created is suppressed.\end{itemize}
In all cases the leptogenesis scale bound obtained in the hierarchical case, Eq.~(\ref{MNbound}), remains essentially unchanged.

For fermion triplets the situation is different, once again due to the gauge scatterings. This depends on the regime we consider:
\begin{itemize}
\item[-] a) \underline{Yukawa regime}: in this regime as explained above the gauge scatterings have no effect in the unflavoured case. However they can have an effect in the flavour case.
To understand this let us consider for example a hierarchical seesaw spectrum with a triplet mass equal to the bound value of Eq.~(\ref{msigmabound}). As already mentioned above, for this value of the mass, the Yukawa regime holds for $\tilde{m}\geq \tilde{m}_A=2.2\cdot 10^{-3}$~eV and the maximum value of the asymmetry (giving the bound of Eq.~(\ref{msigmabound})) is obtained for $\tilde{m}\simeq \tilde{m}_A$.
The gauge scatterings suppress the efficiciency by a factor of $\sim Y^{eq}_\Sigma(z=z_A)/Y^{eq}_\Sigma(z<<1)\sim 10^{-2}$ with $z_A=7.2$ the value of $z$ where $4\gamma_A$ becomes smaller than $\gamma_D$.
Now, as also explained above, for $\tilde{m}\gtrsim \tilde{m}_A$ ($\tilde{m}>>\tilde{m}_A$) the gauge scattering have a moderate (no) effect on the efficiency, not because they are out-of-thermal equilibrium but because they are slower than the decays/inverse decays, $\gamma_D>4\gamma_A$.
Now suppose that playing with flavour we render the inverse decay term harmless. Thus, in this case one has no more suppression from the inverse decay but still one from the (flavour blind) gauge scattering processes \cite{AristizabalSierra:2010mv}, since they are in thermal equilibrium. As a result one understands that for $\tilde{m}\gtrsim \tilde{m}_A$, one can get a larger asymmetry than in the unflavoured case, but still a gauge scattering suppressed result (unless one takes a so large value of $\tilde{m}$ that $z_A\lesssim 1$). For numerical examples of this kind see Fig.~5 of Ref.~\cite{AristizabalSierra:2010mv}.
A consequence of this behavior we want to point out here is that the absolute lower bound on the mass of the decaying triplet for a hierarchical spectrum is essentially the one we get in the type-I case, Eq.~(\ref{MNbound}),
\begin{equation}
m_\Sigma >\,\,\sim 10^9\,\hbox{GeV}\, .
\end{equation}
but this requires large values of $\tilde{m}$ of order at least $\sim 10$~eV. For $\tilde{m}=0.05$~eV one can reduces the bound of Eq.~(\ref{msigmabound}) only by a factor $\sim3$. A same effect could be used in principle for a quasi-degenerate triplet spectrum.

\item[-] b) \underline{Gauge regime}: as explained above, in this regime all the suppression of the asymmetry comes from the (flavour blind) gauge scatterings and the inverse decay term can be neglected. Therefore there are no flavour effects,  as discussed in Ref.~\cite{AristizabalSierra:2010mv}. In particular the lower bound on the leptogenesis scale of Eq.~(\ref{msigmaboundtev}) still holds, because this bound is obtained deeply in the gauge regime. 
The latter statement can be understood from the fact that for example for a one TeV triplet mass, sphalerons decouple at $z\sim 7.5$, and at this temperature $\gamma_D$ is much smaller than $\gamma_A$ (unless $\tilde{m}$ is really large, above $\sim$~KeV).

\end{itemize}


\subsection{Consequences for LHC}

Given the smallness of neutrino masses, a seesaw state
 with a mass accessible at colliders has typically very suppressed Yukawa couplings, far too small to produce them in large enough numbers.
However, a fermion triplet, unlike a right-handed neutrino, can be produced at colliders via gauge interactions, via Drell-Yann pair production. This has been analyzed in Refs.~\cite{Ma:2002pf,Bajc:2006ia,Bajc:2007zf,Franceschini:2008pz,delAguila:2008cj,Arhrib:2009mz,Li:2009mw,Bandyopadhyay:2010wp,Biggio:2011ja}. For a luminosity of 10/fb and 14 TeV energy one should be able to produce at LHC about  
30 ($10^4$) pairs of triplets with mass equal to 1 TeV (250~GeV). There are 3 main types of processes one can look at, see for instance Ref.~\cite{Franceschini:2008pz}: a) $pp\rightarrow \bar{\nu} W^+ W^\pm l^\mp\rightarrow 4 \hbox{jets} +\hbox{missing energy}+ \hbox{a charged lepton}$, the signal with largest rate but largest background too, b) lepton number violating signature, $pp\rightarrow \Sigma^\pm \Sigma^0\rightarrow l_1^\pm l_2^\pm Z W^\mp$, with smaller rate but much smaller background and c) lepton flavour violating processes like $pp \rightarrow l_1\bar{l}_2 4 \hbox{jets}$, with a rate similar to the lepton number violating signals but with larger background. All possibilities are currently under study at LHC and the most promising channels are probably the lepton violating ones. An integrated luminosity of $10/fb$ ($100/fb$) with 14 TeV of energy should allow to see this signal  for $m_\Sigma$ up to $\sim$~0.8 TeV (1.2~TeV).
This is below the absolute leptogenesis scale lower bound of Eq.~(\ref{msigmaboundtev}). To establish the fermion triplet leptogenesis framework at LHC appears consequently hopeless. At the most one could still observe a lighter fermion triplet playing no role for leptogenesis (along the $N_2$ leptogenesis framework mentioned above).
Note that in this case one can take advantage of the fact that at such low scales, neutrino masses typically require small Yukawa couplings, which means slow triplet decay. As a result low seesaw scale models lead  to displaced vertices \cite{Franceschini:2008pz}. One gets $\tau_\Sigma\simeq 0.3 \hbox{mm}\cdot (m_\Sigma/100\,\hbox{GeV})^2\cdot(0.05\,\hbox{eV}/\tilde{m}$). The ATLAS and CMS detectors could in principle detect such kind of displaced vertices as long as they are above $\sim 0.1$~mm. In this way it could also be that a light fermion triplet is observed with Yukawa interactions that would be too large to be compatible with the $N_2$ leptogenesis scenario (i.e.~leading to too much washout of any preexisting L asymmetry). Such an observation could exclude any baryogenesis scenario occuring at a higher scale.

\section{Leptogenesis from the decay of a scalar triplet}


A scalar triplet, as a fermion triplet, undergo gauge interactions which can thermalize it and reduce the efficiency. In addition it has two properties which distinguish it from the fermion seesaw frameworks:
\begin{itemize}
\item A scalar triplet has two totally different types of decay, to a pair of leptons or to a pair of Higgs doublets. At tree level one obtains
\begin{eqnarray}
\Gamma(\Delta_L\rightarrow \bar{L}\bar{L})&=&\frac{m_\Delta}{8 \pi} Tr[Y_\Delta Y_\Delta^\dagger]=B_L \Gamma_\Delta\,,\\
\Gamma(\Delta_L\rightarrow {H}{H})&=&\frac{1}{8 \pi} \frac{|\mu_\Delta|^2}{m_\Delta}=B_H \Gamma_\Delta\,,
\end{eqnarray}
with $\Gamma_\Delta$ the total decay width of the scalar triplet, and $B_{L,H}$ the branching ratios ($B_L+B_H=1$).
At one loop, from the definition of the CP asymmetry in Table.~1, and taking into account CPT symmetry one gets
\begin{eqnarray}
\Gamma(\bar{\Delta}_L\rightarrow {L}{L})&=& \Gamma_\Delta(B_L+\varepsilon_\Delta/2)\,,\label{deltadecay1}\\
\Gamma(\Delta_L\rightarrow \bar{L}\bar{L})&=& \Gamma_\Delta(B_L-\varepsilon_\Delta/2)\,,\\
\Gamma(\bar{\Delta}_L\rightarrow \bar{H}\bar{H})&=& \Gamma_\Delta(B_H-\varepsilon_\Delta/2)\,,\\
\Gamma({\Delta}_L\rightarrow {H}{H})&=& \Gamma_\Delta(B_H+\varepsilon_\Delta/2)\,.\label{deltadecay4}
\end{eqnarray}
The fact that the scalar triplet has two different decays and that lepton number is violated only if both types of decay coexist has important consequences for the efficiency. As we will see below this allows to have no suppression of the efficiency, even at low scale where the gauge scatterings are very fast.\footnote{Actually in the type-I and III framework one can also have a similar structure with similar properties \cite{Blanchet:2009kk} if one assumes an approximately conserved $L$ symmetry, in case one has two types of Yukawa couplings, L-breaking and L-violating. For the scalar triplet this is generic.}
\item A scalar triplet is not a self-conjugate state. Therefore a triplet-antitriplet asymmetry can be generated, leading to an extra Boltzmann equation, see below. Actually for a fermion triplet one could believe at first sight that an asymmetry could be generated between the charged triplets of opposite signs. However this is not the case, at least in the $SU(2)_L$ symmetric limit, since in this limit
one should get the same result than for the neutral triplet component.
The latter cannot have any $\Sigma^0$-$\bar{\Sigma}^0$ asymmetry since it is a Majorana fermion. More explicitly in this $SU(2)_L$ limit the equalities $\Gamma(\Sigma^+\rightarrow \nu \phi^+)=\Gamma(\Sigma^-\rightarrow l \phi^0)$ and $\Gamma(\bar{\Sigma}^+\rightarrow \bar{\nu} \phi^-)=\Gamma(\bar{\Sigma}^-\rightarrow \bar{l} \phi^{0*})$ hold. Therefore if one call these 2 decay widths as "a" and "b" respectively, lepton number is violated because $a\neq b$ but the number of positively charged triplets remains equal to the number of negatively charged triplets (simply because both components change by the same $\propto (a+b)$ amount). And there is no $\Sigma^+$-$\Sigma^{-c}$ asymmetry either, since these two components mix through their Dirac mass. 
Only for the scalar triplet can we have an "unprotected" particle-antiparticle asymmetry for the seesaw states. The extra Boltzmann equation it implies has important consequences, see below.
\end{itemize} 

Let us start with a few general formulas which apply for a scalar triplet.
The lepton asymmetry produced by the decay of a heavy triplet is given by the general formula
\begin{equation}
\frac{n_L}{s}=\varepsilon_\Delta \eta \frac{n_\Delta+n_{\bar{\Delta}}}{s}\Big|_{T>>m_\Delta}\,,
\end{equation}
so that the baryon asymmetry produced is
\begin{equation}
\frac{n_B}{s}= 3 \frac{135 \zeta(3)}{4\pi^4 g_\star}C_{L\rightarrow B} \varepsilon_\Delta \eta=-0.0041 \varepsilon_\Delta \eta \,.
\label{nBoversscalar}
\end{equation}
The factor of 3 in the last equation accounts for the fact that each triplet component produces an asymmetry $\varepsilon_\Delta$. 
Again the $\eta$ efficiency factor is defined in such a way that it takes the value unity if all decays occur out-of-equilibrium and if there is no wash-out effect from $\Delta L\neq0$ processes. 

The total decay width of a scalar triplet can be written as
\begin{equation}
\Gamma_\Delta=\frac{1}{8\pi} \frac{m_\Delta^2\tilde{m}_\Delta}{v^2}\frac{B_L+B_H}{\sqrt{B_LB_H}}\,,
\label{Gammatriplet}
\end{equation}
where 
\begin{equation}
\tilde{m}_\Delta^2\equiv {\rm Tr}[\, {\cal M}_\nu^{\Delta\dagger}{\cal M}_\nu^\Delta]=\sum_i (m_{\nu_i}^{\Delta })^2=Tr[Y_\Delta^\dagger Y_\Delta] |\mu_\Delta|^2\frac{v^4}{m_\Delta^4}\quad (i=1,2,3)\,,
\end{equation}
with $m_{\nu_i}^{\Delta}$ the eigenvalues of the neutrino mass matrix induced by the scalar triplet. As a result the following bound applies
\begin{equation}
\Gamma_\Delta\geq \frac{1}{4\pi} \frac{m_\Delta^2\tilde{m}_\Delta}{v^2}\,.
\label{boundGammatriplet}
\end{equation}
This bound is saturated for $B_L=B_H=1/2$. 
Eq.~(\ref{Gammatriplet}) divided by the Hubble constant can be written as
\begin{equation}
\frac{\Gamma_\Delta}{H}\Big|_{T=m_\Delta}=\frac{\sqrt{\sum (m_{\nu_i}^\Delta)^2}}{0.001\,\hbox{eV}}\frac{1}{2}\frac{B_L+B_H}{\sqrt{B_LB_H}}\,.
\label{Gammatripletbis}
\end{equation}
It means that if the triplet contributes for $\nu$ masses above the $\sim10^{-3}$~eV level, the triplet decays anyway in thermal equilibrium. It shows also that for fixed contribution to the neutrino masses, the most we are off of the $B_L=B_H$ case, the most the triplet is in thermal equilibrium. In this case the triplet decay/inverse decay processes can be in thermal equilibrium even if this triplet contributes very little to the neutrino masses. The reason why one gets the sum of the neutrino mass squared, rather than the lightest neutrino mass as for the type-I and III cases, Eq.~(\ref{mtildemintypeI}), stems from the fact that ${\cal M}_\nu^\Delta$ could be any generic matrix, while ${\cal M}_\nu^N$ has rank 1.

For a hierarchical spectrum of heavy states, the CP asymmetry given in Eq.~(\ref{epsIIhierarc}) is bounded from above. Rewriting it as
\begin{equation} 
\varepsilon_\Delta  =  - \frac{1}{2\pi}\frac{m_\Delta}{v^2}\sqrt{B_L B_H}
 \frac{{\rm Im}\,{\rm Tr}\, [{\cal M}_\nu^\Delta {\cal M}_\nu^{H\dagger} ]}{\tilde{m}_\Delta}\,,
 \label{epsDelta}
 \end{equation}
one gets the bound \cite{Hambye:2005tk}
 \begin{equation}
  | \varepsilon_\Delta |  \le 
 \frac{1}{2\pi} \frac{m_\Delta}{v^2}\sqrt{B_L B_H\sum_i  m_{\nu_i}^2} \,,
 \label{epsboundDelta}
 \end{equation}
which holds for the decay of a heavy triplet, no matter what is the nature of the heavy seesaw states. 
The bound of Eq.~(\ref{epsboundDelta}) is a model independent bound. If extra information on how the neutrino mass matrix decomposes as the sum of ${\cal M}_\nu^\Delta$ and ${\cal M}_\nu^H$ is known (as e.g. in particular neutrino mass models) the bound of Eq.~(\ref{epsboundDelta}) can be strengthened since one gets in this case
 \begin{equation}
  | \varepsilon_\Delta |  \le 
 \frac{1}{2\pi} \frac{m_\Delta}{v^2}\sqrt{B_L B_H\hbox{Min}[\sum_i  m_{\nu_i}^2,\tilde{m}_H^2]} \,,
 \label{epsboundDelta2}
 \end{equation}   
 with $\tilde{m}_H^2\equiv Tr[{\cal M}_\nu^{H\dagger} {\cal M}_\nu^H ]$.
Here too the reason why one gets the sum of the neutrino mass squared, rather than $m_{\nu_3}-m_{\nu_1}$ as in Eq.~(\ref{epsIbound}), is associated to the fact that ${\cal M}_\nu^\Delta$ could be any generic matrix, while ${\cal M}_\nu^N$ has rank 1. This bound implies a lower bound on $m_\Delta$ which we will determine from the Boltzmann equation below. Strictly speaking, Eqs.~(\ref{epsboundDelta})-(\ref{epsboundDelta2}) are valid only in the hierarchical limit. In the general case there exists also a "perturbativity" bound on the CP-asymmetry, one obtains by requiring that the one loop contribution to the various decay widths doesn't exceed the tree level one,
\begin{equation}
\varepsilon_\Delta\le2\,\hbox{Min}(B_L,B_H)\,.
\label{epsboundscalarunit}
\end{equation}

\subsection{Efficiency of a decaying scalar triplet}

As explained above, if $L$, $C$ and $CP$ are violated there is no symmetry principle which could prevent that a triplet-antitriplet asymmetry, $\Delta_T\equiv Y_\Delta-Y_{\bar{\Delta}}$, would be created when the triplets decay. As a result there are three independent coupled Boltzmann equations instead of two. The set of equations one gets is 
 \begin{eqnarray}
sHz \frac{d\Sigma_{T}}{dz} &=&
  -\bigg(\frac{\Sigma_{T}}{\Sigma_{T}^{\rm eq}}-1\bigg)\gamma_D
  -2\bigg(\frac{\Sigma_{T}^2}{\Sigma_{T}^{2\rm eq}}-1\bigg)\gamma_A \,, 
\label{BoltzST}\\
sHz \frac{d\Delta_L}{dz} &=&  \gamma_D \varepsilon_L \bigg(\frac{\Sigma_{T}}{\Sigma_{T}^{\rm eq}}-1\bigg)
-2\gamma_DB_L(\frac{\Delta_L}{Y_L^{\rm eq}}+\frac{\Delta_T}{\Sigma_T^{\rm eq}}) +X\,,
\label{BoltzdL}\\
sHz \frac{d\Delta_H}{dz} &=& \gamma_D \varepsilon_L \bigg(\frac{\Sigma_{T}}{\Sigma_{T}^{\rm eq}}-1\bigg)
-2\gamma_DB_H(\frac{\Delta_H}{Y_H^{\rm eq}}-\frac{\Delta_T}{\Sigma_T^{\rm eq}})+X\,,
\label{BoltzdH}
\\
sHz \frac{d\Delta_T}{dz} &=&-\gamma_D\left(\frac{\Delta_T}{\Sigma_{T}^{\rm eq}}+B_L
\frac{\Delta_L}{Y_{L}^{\rm eq}}-B_H \frac{\Delta_H}{Y_{H}^{\rm eq}}\right) ,
 \label{BoltzdT}
\end{eqnarray}
with $z\equiv m_\Delta/T$, $\Sigma_T=(n_\Delta+n_{\bar{\Delta}})/s$, $\Delta_L=(n_L-n_{\bar{L}})/s$, $\Delta_H=(n_H-n_{\bar{H}})/s$ and
\begin{equation}X=  -2
(\frac{\Delta_L}{Y_L^{\rm eq}} + \frac{\Delta_H}{Y_H^{\rm eq}}) 
(\gamma_{Ts}^{\rm sub}+\gamma_{Tt})\,.
\end{equation}
Out of these 4 equations, only three are independent, due to the decay sum rule $2\Delta_T+\Delta_H-\Delta_L=0$, which follows from the fact that a $\Delta_L$ triplet can decay  to $HH$ or $\bar{L}\bar{L}$ but not to their conjugated states, Eqs.~(\ref{deltadecay1})-(\ref{deltadecay4}). The decay/inverse decay reaction rate is 
$\gamma_D=(n_\Delta^{eq}+n_{\bar{\Delta}}^{eq})\Gamma_\Delta K_1(z)/K_2(z)$. $\gamma_A$ comes from all gauge scattering processes of the triplets, $\delta \delta'\leftrightarrow, GG',\,f \bar{f},\, H\bar{H}$ with $G^{(')}$ all possible gauge bosons. 
One has 
\cite{Cirelli:2007xd,Hambye:2005tk}
\begin{eqnarray}
\hat\sigma_A= 
\frac{6}{72 \pi}&\Big[&(15 C_1 - 3 C_2) \beta +(5C_2-11C_1)\beta^3\nonumber\\
&&+3(\beta^2-1)[2C_1+C_2(\beta^2-1)]\ln \frac{1+\beta}
{1-\beta}\Big]\,,
\end{eqnarray}
where  $x=s/m^2_\Delta$, $\beta= \sqrt{1-4/x}$, $C_1=3g^4/2+3 g_Y^4+12 g^2g_Y^2$ and $C_2=6g^4+3 g_Y^4+12 g^2g_Y^2$.
The other reaction rates parametrize the effects of $\Delta L=2$ scatterings, $LL\leftrightarrow \bar{H}\bar{H}$ and $L\bar{H}\leftrightarrow \bar{L}{H}$. Their analytical expressions can be found in Ref.~\cite{Hambye:2005tk}.

To understand the structure of the Boltzmann equations it is useful to proceed step by step, in a way similar to the discussion of the fermion triplet above:
\begin{itemize}
\item[(a)] To start let us consider only the decay/inverse decay term without washout from inverse decays and without scattering processes, i.e.~consider only the first term of the first 3 Boltzmann equations, Eqs.~(\ref{BoltzST})-(\ref{BoltzdH}). In this case no matter is $\gamma_D$ the efficiency is unity. If $\gamma_D/(n_\Delta^{eq} H)>>1$, i.e.~$\Gamma_\Delta>>H$, the decay/inverse decay processes are in deep thermal equilibrium,  i.e.~$\Sigma_T/\Sigma_T^{eq}-1<<<1$, but this gives the same result as the opposite case ($\Gamma_\Delta<<H$) because all equations involve the same $\gamma_D (\Sigma_T/\Sigma_T^{eq}-1)$ product, which remains constant. 
\item[(b)] Now let us add the $\gamma_A$ gauge term in Eq.~(\ref{BoltzST}). As for a fermion triplet $\gamma_A/n_\Delta^{eq}H$ scales as $m_{Pl}/T$ for $T>>m_\Delta$. For $T<<m_\Delta$ it is a good approximation to consider just the s-wave contribution which gives Eqs.~(\ref{hatsigmasp})-(\ref{gammasigmasp}) with $m_\Sigma$ replaced by $m_\Delta$ and
\begin{equation}
c_s=\frac{9g^4+12g^2g_Y^2+3g_Y^4}{2 \pi}\,.
\end{equation}
This gives a doubly Boltzmann suppressed reaction rate, $\gamma_A\propto e^{-2m_\Delta/T}$, since it involves 2 external heavy particles,  so that the thermalization rate involves only one Boltzmann suppression power,
$\gamma_A/n_\Delta^{eq}H\propto e^{-m_\Delta/T} m_{Pl}/\sqrt{m_\Delta T}$. The maximum value of $\gamma_A/n_\Delta^{eq}H$ is reached for $T\sim m_\Delta$, where it takes the value   
\begin{equation}
\frac{\gamma_A}{n_\Delta^{eq}H}\sim 7 \cdot \frac{10^{13}\,\hbox{GeV}}{m_\Delta}\,.
\end{equation}
Therefore, similarly to the fermion triplet case, as soon as $m_\Delta$ is below $\sim 10^{14}$~GeV,  $\gamma_{A}/(n_\Delta^{eq} H)>1$ at $T\sim m_\Delta$,  so that $\Sigma_\Delta \simeq \Sigma_\Delta^{eq}$ at this temperature, and the asymmetry produced is independent of the initial scalar triplet population we start with.  And the smaller is $m_\Delta$ the more the triplet is thermalized at a temperature around its mass. The discussion is the same as for the fermion triplet above.
Above
$m_\Delta\sim10^{14}$~GeV, this gives an efficiency equal to unity. Below this value,
if  for any $T$ where $\gamma_{A}/(n_\Delta H)>1$, the $4 \gamma_A\leq \gamma_D$ inequality holds, the gauge scattering term can be neglected in Eq.~(\ref{BoltzST}) and we are back to case (a) above with efficiency unity, even if the gauge scattering term are in thermal equilibrium for some time. The triplets decay before they get the opportunity to annihilate through gauge interactions. If on the contrary $4 \gamma_A\geq \gamma_D$ down to a temperature $T_A$ below $\sim m_\Delta$, the $\gamma_D$ term can be neglected in Eq.~(\ref{BoltzST}), and $(\Sigma_T/\Sigma_T^{eq}-1)$ (and hence the production of an asymmetry) is suppressed  down to this temperature.
This gives, similarly to Eq.~(\ref{YLsigma2}),
\begin{equation}
Y_L\simeq \varepsilon_\Delta Y_\Delta^{eq}(z_A) (z_A/4+1)\,,
\label{YLscalartriplet2b}
\end{equation}
with $z_A\equiv m_\Delta/T_A$, the value of $z$ where $4 \gamma_A$ gets below $\gamma_D$ (remember that $\gamma_A$ involves one Boltzmann suppression power more than $\gamma_D$).

\item[(c)] Now let us add the $X$ term, i.e.~the $\Delta L=2$ scattering terms. As for the fermion triplet case these terms are important only for masses above $\sim 10^{12-14}$~GeV.
\item[(d)] Finally let us add the wash-out term from the inverse decay, i.e.~all remaining terms  in Eqs.~(\ref{BoltzdL})-(\ref{BoltzdT}), which are all proportional to $\gamma_D$. 

If $B_L\simeq B_H$ the discussion of the $\gamma_A$-$\gamma_D$ interplay is just the same as for the fermion triplet. This can be seen easily from the fact that if $B_L=B_H$, Eq.~(\ref{BoltzdT}) vanishes and one gets a set of 2 independent Boltzmann equations similar to the fermion triplet case. Just as for a fermion triplet one can determine a critical value of $\tilde{m}_\Delta$ below (above) which one lies in the 
"gauge" ("Yukawa") regime.  
This value is given in Fig.~\ref{regimesscal} as a function of $m_\Delta$. For $B_L=B_H=1/2$ it gives about the same efficiency than for a fermion triplet, that is to say about the same efficiency than for the type-I case in the Yukawa regime and an efficiency given by Eq.~(\ref{YLscalartriplet2b}) in the gauge regime. 

\begin{figure}[!t]
\centering
\includegraphics[height=7.5cm]{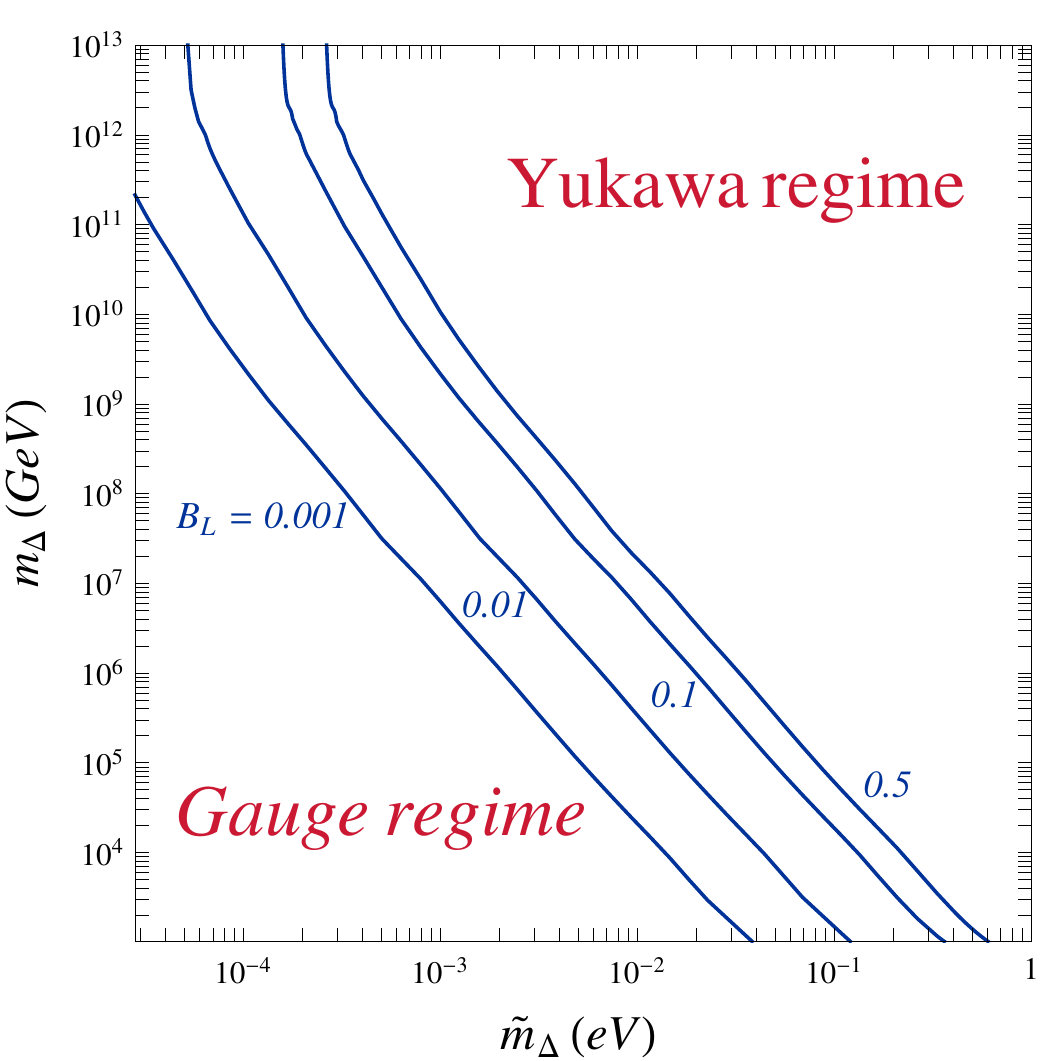}\\
\caption{Gauge versus Yukawa regimes for various values of $B_L$, as a function of $\tilde{m}_\Delta$ and $m_\Delta$. As a function of $B_H$ the separation between both regimes is the same, simply replacing $B_L$ by $B_H$.}  
\label{regimesscal}
\end{figure}

If instead $B_L>>B_H$ or $B_H>>B_L$ the situation is different. In this case it turns out that one gets larger efficiencies. One can even get an efficiency of order unity, even if the triplet mass is orders of magnitudes below $10^{14}$~GeV.
To understand this feature let us consider a value of $\tilde{m}_\Delta$ large enough to insure that the triplet decay before it annihilates for any value of $z\gtrsim 1$. In this case, even if the decay/inverse decay $\gamma_D$ term is deeply in thermal equilibrium, one doesn't get necessarily a suppression of the efficiency.
The point is that in the type-II seesaw model, lepton number violation is, as said above, due to the coexistence of scalar triplet couplings to a pair of lepton and to a pair of Higgs doublet. Therefore, even if the total decay rate is much larger than the Hubble rate, if $B_H <<1$ or $B_L<<1$ (so that 
either the decay to Higgs doublets or the decay rate to leptons is out of thermal equilibrium), lepton number is not washed-out by the inverse decay. In this case the scalar triplet has effectively lepton number 2 or 0 respectively, and lepton number is broken only by the subleading decay process which is out-of thermal equilibrium. In other words, even if the gauge scatterings and the total decay rate are in thermal equilibrium, there is no suppression from the $\gamma_A$ gauge scattering term if $\gamma_D>4 \gamma_A$ 
and there is no suppression either from $\gamma_D$ if one of the decay rate is much smaller than the other one (so that it is out of thermal equilibrium). The third Sakharov condition, i.e.~that "the creation of an asymmetry requires that the decaying particles are out-of-equilibrium", is therefore not exact, if taken literally. The decaying particle can be deeply in thermal equilibrium with the Universe thermal bath without that any washout is induced. What is precisely required is that the inverse decays that are in thermal equilibrium do not break lepton number. 

Note that this is possible only because one has 3 Boltzmann equations for the asymmetries. If there were only two, for $\Delta_L$ and $\Delta_H$, the sum rule $\Delta_H-\Delta_L=0$ (which holds in type-I and type-III leptogenesis) would necessarily lead to a suppression of both asymmetries. 

\begin{figure}[!t]
\centering
\includegraphics[height=7.5cm]{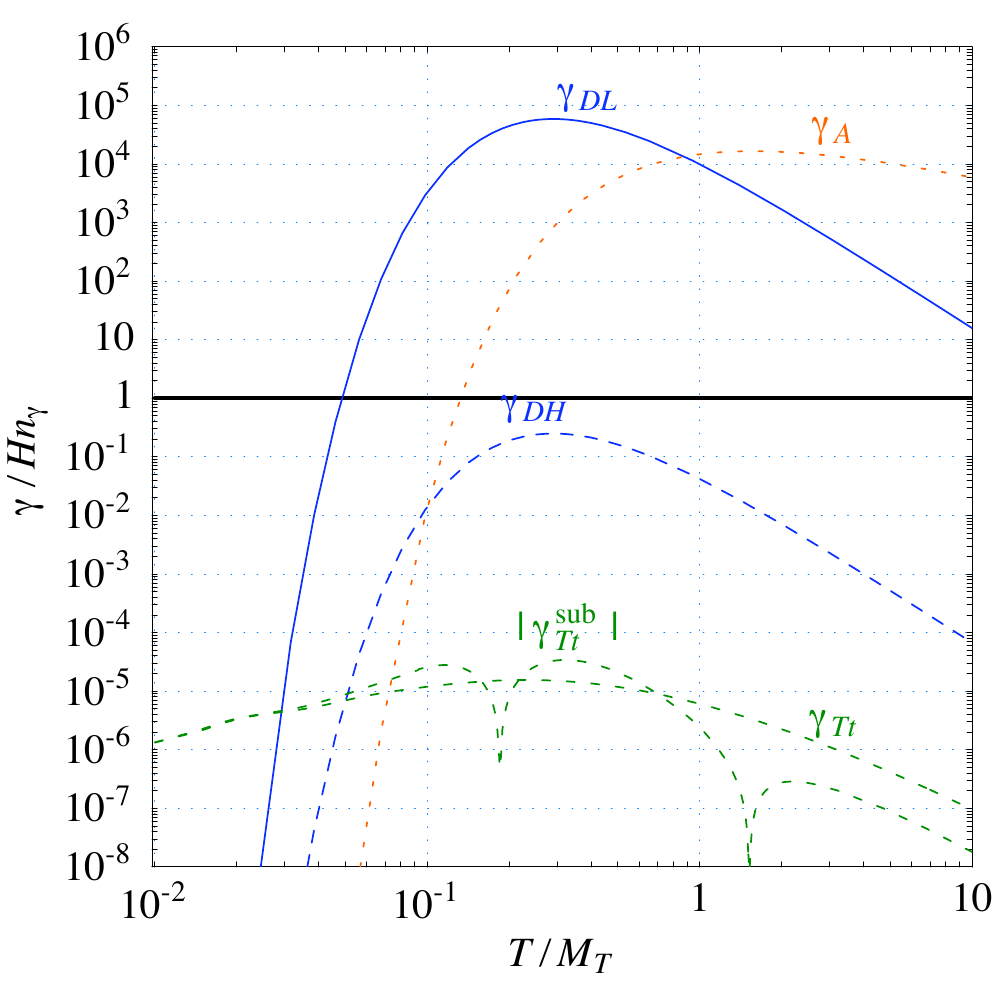}\includegraphics[height=7.5cm]{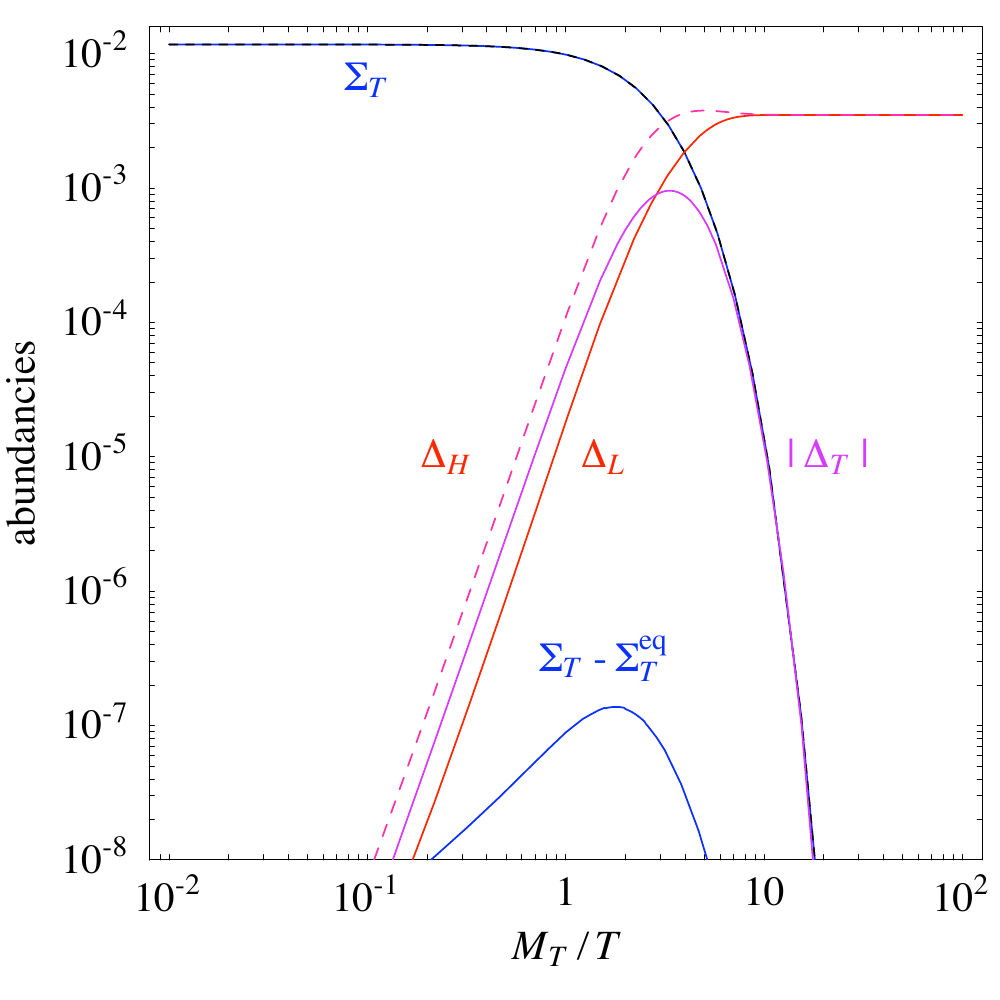}\\
\caption{From Ref.~\cite{Hambye:2005tk}, thermalization rates $\gamma_i/H n_\gamma$ (left panel) and corresponding asymmetry evolutions (right panel) for an example case where $m_\Delta=10^{10}$~GeV, $\tilde{m}_\Delta=0.05$~eV, $\lambda_L=0.1$ giving $B_H=3\cdot 10^{-6}$ and efficiency $\eta\simeq 0.38$.}  
\label{figasymevolscal}
\end{figure}

To see better how effectively a large asymmetry can develop itself in such a case, in Fig.~\ref{figasymevolscal} we give the rates and asymmetry evolution we get for an example of situation where the gauge scatterings as well as the total decay rate are much larger than the Hubble rate, with $B_H<<1$ (so that the decay to Higgs doublets is never in thermal equilibrium). 
This example shows that the first asymmetry to become large is the $\Delta_H$ one, because, if $B_H<<1$,  the washout term proportional to $\gamma_D B_H$ in Eq.~(\ref{BoltzdH}) is ineffective. The Higgs asymmetry produced is not suppressed and increases rapidly. One gets approximately $\Delta_H(T)\simeq \varepsilon_\Delta (\Sigma_T^{eq}(T>>m_\Delta)-\Sigma_T^{eq}(T))$. 
Due to the sum rule $2\Delta_T+\Delta_H -\Delta_L=0$ an equally large  $2\Delta_T-\Delta_L$ develops itself. 
Note that since in this case the scalar triplet has effectively $L=-2$, the $2\Delta_T-\Delta_L$ asymmetry is nothing else than the effective lepton asymmetry. This combination is not washed-out since there is no fast decay breaking this effective lepton number. On the other hand the $\Delta_L$ asymmetry is washed-out largely by the inverse decay term proportional to $\gamma_D B_L$ in Eq.~(\ref{BoltzdL}), since this term is large. This means that in a first step $2 \Delta_T-\Delta_L\simeq 2 \Delta_T$. The lepton number asymmetry is therefore stored in the triplet asymmetry. Finally, later on, as all triplets will ultimately decay (to leptons mostly since $B_L>>B_H$), all the effective lepton asymmetry stored in the triplet asymmetry will be transferred to $\Delta_L$, i.e.~to a lepton asymmetry in terms of leptons. Or in other words, since the $\Delta_H$ asymmetry produced is anyway large since it is not affected by any washout term, and since ultimately, after all triplets have decayed, the sum rule $\Delta_H -\Delta_L=0$ applies, all the $2 \Delta_T-\Delta_L$ effective lepton asymmetry is necessarily transferred to the $\Delta_L$ asymmetry. This pattern can be clearly seen in Fig.~(\ref{figasymevolscal}).
\end{itemize}

\begin{figure}[!t]
\centering
\includegraphics[height=7.5cm]{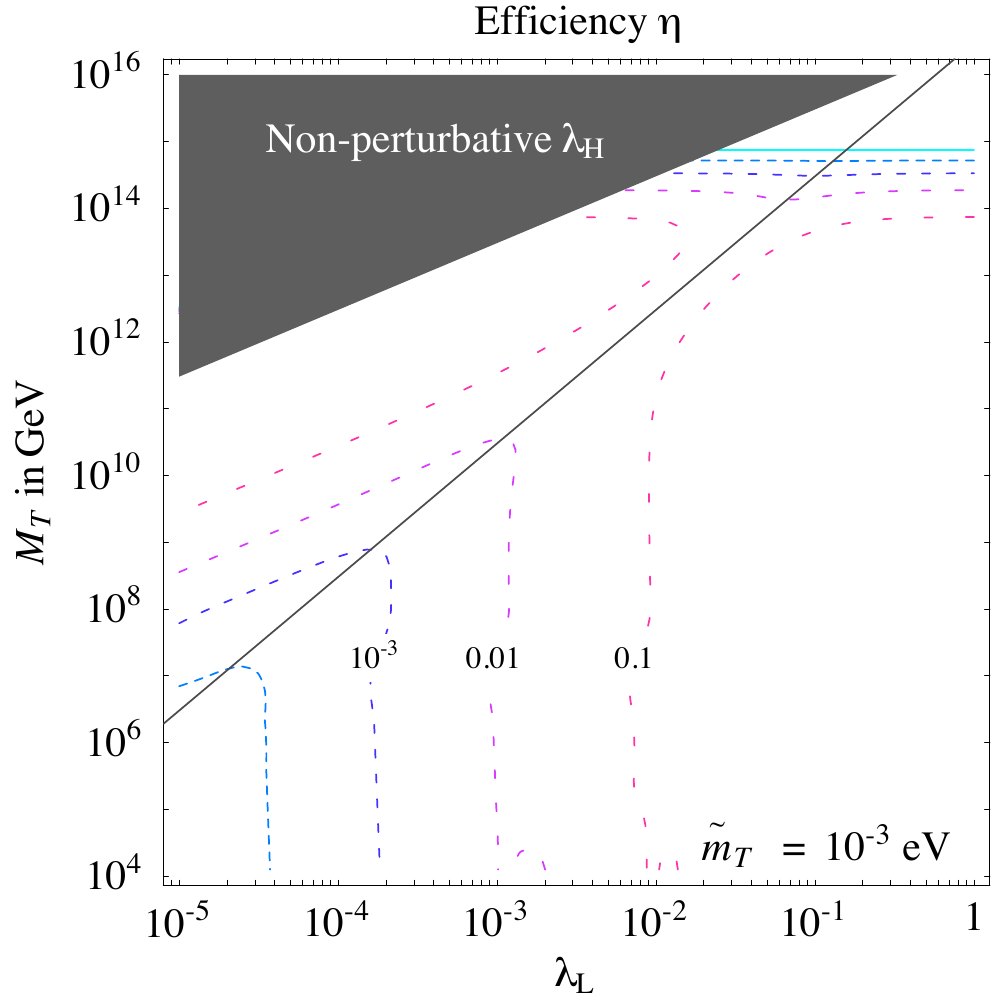}\includegraphics[height=7.5cm]{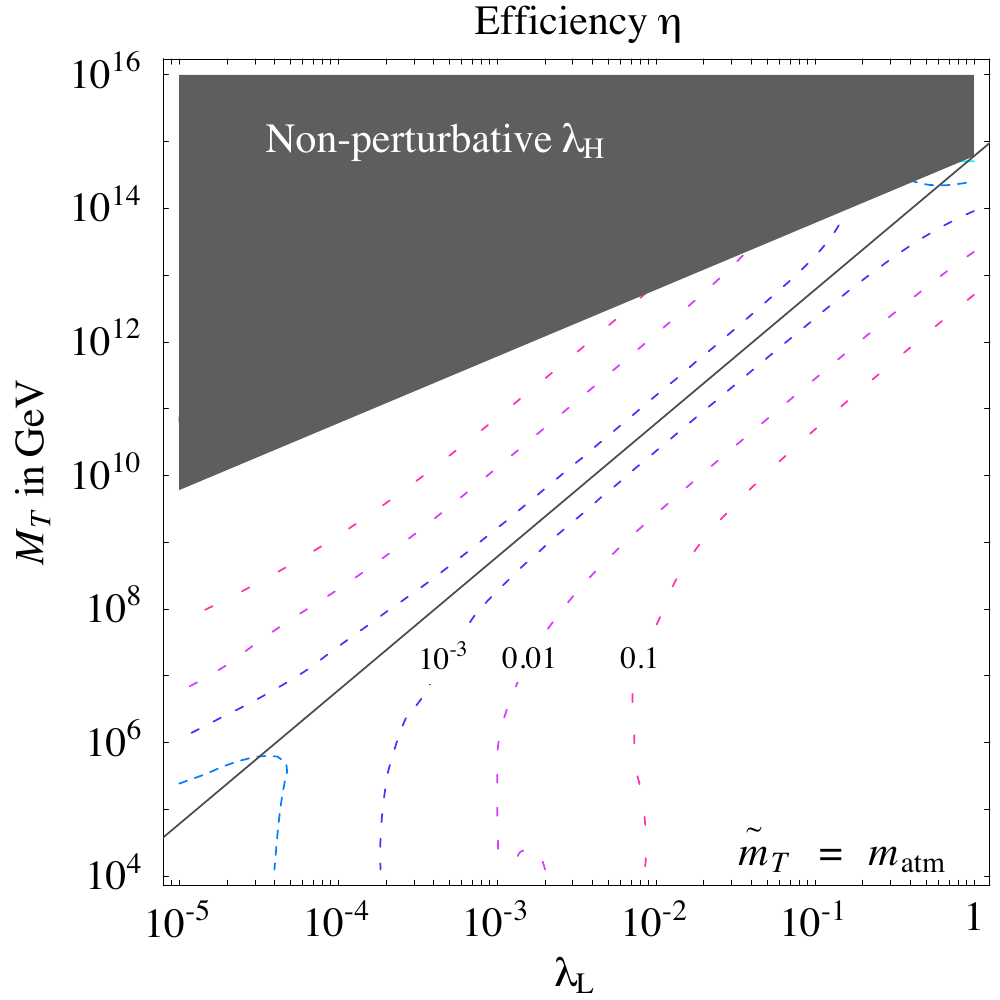}\\
\caption{From Ref.~\cite{Hambye:2005tk}, scalar triplet efficiency obtained as a function of $\lambda_L$ and $m_\Delta$ for $\tilde{m}_\Delta= 10^{-3}$~eV (left panel) and $\tilde{m}_\Delta=0.05$~eV (right panel). The diagonal line corresponds to the $B_L=B_H=1/2$ case. The shaded regions are regions where $\lambda_H\equiv \sqrt{2} \mu/m_\Delta$ is larger than one.}  
\label{etascalinterm}
\end{figure}
\begin{figure}[!t]
\centering
\includegraphics[height=7.5cm]{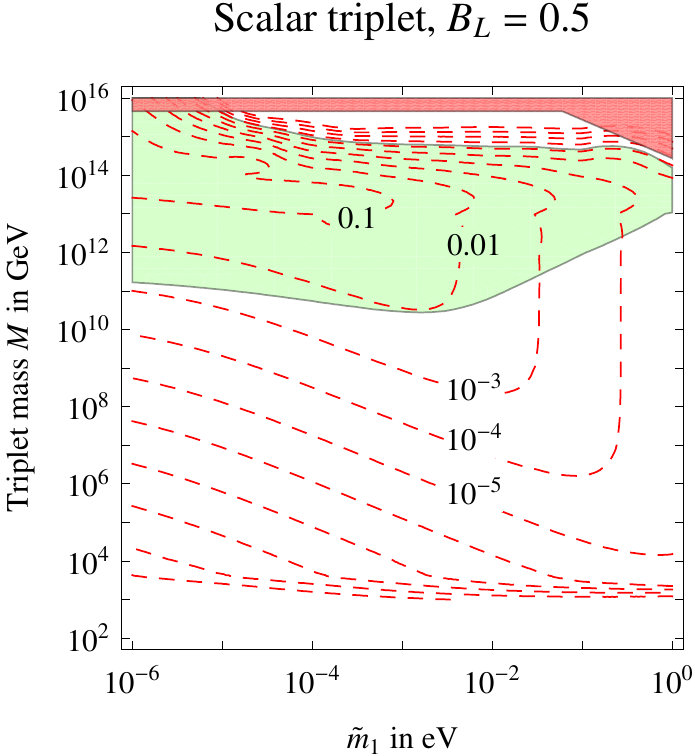}\\
\caption{Scalar triplet efficiency obtained in the case $B_L=B_H=1/2$, as a function of $\tilde{m}_\Delta$ and $m_\Delta$, from Ref.~\cite{Strumia:2008cf}. Similar results can also be found in Refs.~\cite{Hambye:2005tk,Chun:2006sp}}  
\label{etastrumiascal}
\end{figure}

From all these considerations one can understand the behavior of the efficiency one obtains for a scalar triplet, Figs.~\ref{etascalinterm} and \ref{etastrumiascal}.
The efficiency depends on 3 parameters, $m_\Delta$, $\Gamma_\Delta$ and $B_L$. Equivalently, as has been done in these figures \cite{Hambye:2005tk,Strumia:2008cf}, it may be expressed as a function of $m_\Delta$, 
$\tilde{m}_\Delta$ and $\lambda_L$ with $\lambda_L^2\equiv 2 Tr[Y_\Delta Y_\Delta^\dagger]$, so that $\Gamma_\Delta=(m_\Delta/16\pi)(\lambda_L^2+4\tilde{m}_\Delta^2m_\Delta^2/(\lambda_L^2v^4))$ and $B_L=\lambda_L^2 m_\Delta/(16 \pi\Gamma_\Delta)$.
Fig.~\ref{etascalinterm} shows how the efficiency varies as a function of $m_\Delta$ and $\lambda_L$ for $\tilde{m}_\Delta=0.05$~eV and $\tilde{m}_\Delta=10^{-3}$~eV. The first case holds in particular for the situation where the decaying triplet dominates neutrino masses with a hierarchical spectrum of light neutrinos. The second situation corresponds to a subleading  
triplet contribution to neutrino masses (with leading contribution coming from the heavier seesaw states). The efficiency is also given in Fig.~\ref{etastrumiascal} as a function of $\tilde{m}_\Delta$ for $B_L=B_H=1/2$. Let us first consider this case.

\underline{$B_L=B_H=1/2$ case}: as anticipated the efficiency displayed in Figs.~\ref{etascalinterm} and \ref{etastrumiascal} has essentially the same properties than for the fermion triplet, see Section 3. Both efficiencies do not differ by more than a factor of a few, compare for example Figs.~\ref{etastrumiascal} and \ref{figeffic}. For $\tilde{m}_\Delta=10^{-3}$~eV one always lies in the "gauge regime", unless $m_\Delta\gtrsim 10^{11}$~GeV, whereas, for $\tilde{m}=0.05$~eV, one always lies in the "Yukawa regime", unless $m_\Delta$ is below $\sim 10^6$~GeV. Remember the transition line between the gauge and Yukawa regimes for $B_L=B_H=1/2$ is given in Fig.~\ref{regimesscal}.
Therefore for $\tilde{m}_\Delta=0.05$~eV  and unless $m_\Delta<10^6$~GeV, the efficiency one gets is similar to the one one gets in the type-1 case. In first approximation the efficiency scales as $\sim1/\tilde{m}_\Delta$, almost independently of $m_\Delta$.
For $\tilde{m}_\Delta=10^{-3}$~eV, on the other hand, one gets only a very mild suppression from inverse decay at high mass but one do get a suppression from gauge scatterings. This gives an efficiency  which scales as $\sim m_\Delta$ for $m_\Delta$ below $10^{11}$~GeV. $\Delta L=2$ scatterings induce a suppression only for $m_\Delta> 10^{14}$~GeV.

\underline{$B_L\neq B_H$ case}: as expected above, Figs.~\ref{etascalinterm} and \ref{etastrumiascal} display an efficiency which increases off the $B_L=B_H$ diagonal. Note that  Eq.~(\ref{Gammatriplet}) can be rewritten as
\begin{equation}
\frac{\Gamma_\Delta}{H(T=m_\Delta)}= \Big(\frac{\tilde{m}_\Delta}{0.0012\,\hbox{eV}}\Big) \cdot \frac{1}{2}\cdot \frac{B_L+B_H}{\sqrt{B_LB_H}}\,,
\end{equation}
which means that the total inverse decay rate is in thermal equilibrium  at $T\sim m_\Delta$ provided $\tilde{m}_\Delta=\sqrt{(m_{\nu_i}^{\Delta})^2}\gtrsim 0.002\,\hbox{eV} \cdot  \frac{\sqrt{B_LB_H}}{B_L+B_H}$. Therefore,
for fixed value of $\tilde{m}_\Delta$ and $m_\Delta$, as one increases $B_L$ or $B_H$, the total decay rate increases, so that $\gamma_D/\gamma_A$ increases (suppressing the gauge effect) and at the same time $L$ gets less and less broken as $B_H \gamma_D$ or $B_L \gamma_D$ decreases (suppressing the inverse decay effect), see Eq.~(\ref{Gammatriplet}).  
For $\tilde{m}_\Delta=0.05$~eV this is the second effect which is mostly responsible for the increase of the efficiency off the diagonal line in Fig.~\ref{etastrumiascal}. For $\tilde{m}_\Delta=10^{-3}$~eV this is mostly the first one. In fact for $B_L\neq B_H$ one gets out of the gauge regime (i.e. gauge effects cease to be relevant, see Section 3) as soon as 
\begin{equation}
\tilde{m}_\Delta < \tilde{m}_\Delta^A\cdot  \frac{\sqrt{B_LB_H}}{B_L+B_H} \cdot 2\,,
\end{equation}
with $\tilde{m}^A_\Delta$ the transition value one gets for $B_L=B_H=1/2$, see Fig.~\ref{regimesscal}. In this figure one can find, for various values of $B_L$, the values of $\tilde{m}_A$ separating both regimes.
Similarly there is no sizeable inverse decay effect as soon as  
\begin{equation}
Min[B_L,B_H]\cdot \frac{\Gamma_\Delta}{H(T=m_\Delta)}\lesssim 1 \,.
\end{equation}
For $B_L$ smaller (larger) than $B_H$, this inequality can be rewritten as $\lambda_L^2/m_\Delta<17\cdot16 \pi/M_{Pl}$ ($\mu_\Delta^2/m_\Delta^3<17\cdot8 \pi/M_{Pl}$). 
In the Yukawa regime, when $K_{Min}\equiv Min[B_L,B_H]\Gamma_\Delta/H|_{T=m_\Delta}$ is larger than unity, the efficiency due to the inverse decay processes scales in first approximation as $1/K_{Min}$, that is to say as $(\lambda_L^2/m_\Delta)^{-1}$ and $(\mu_\Delta^2/m_\Delta^3)^{-1}=  (\lambda_L^2/m_\Delta)(v^4/\tilde{m}_\Delta^2)/2$ for $B_L<<1$ and $B_H<<1$ respectively.

\subsection{Baryon asymmetry from a decaying scalar triplet}

\begin{figure}[!t]
\centering
\includegraphics[height=7.5cm]{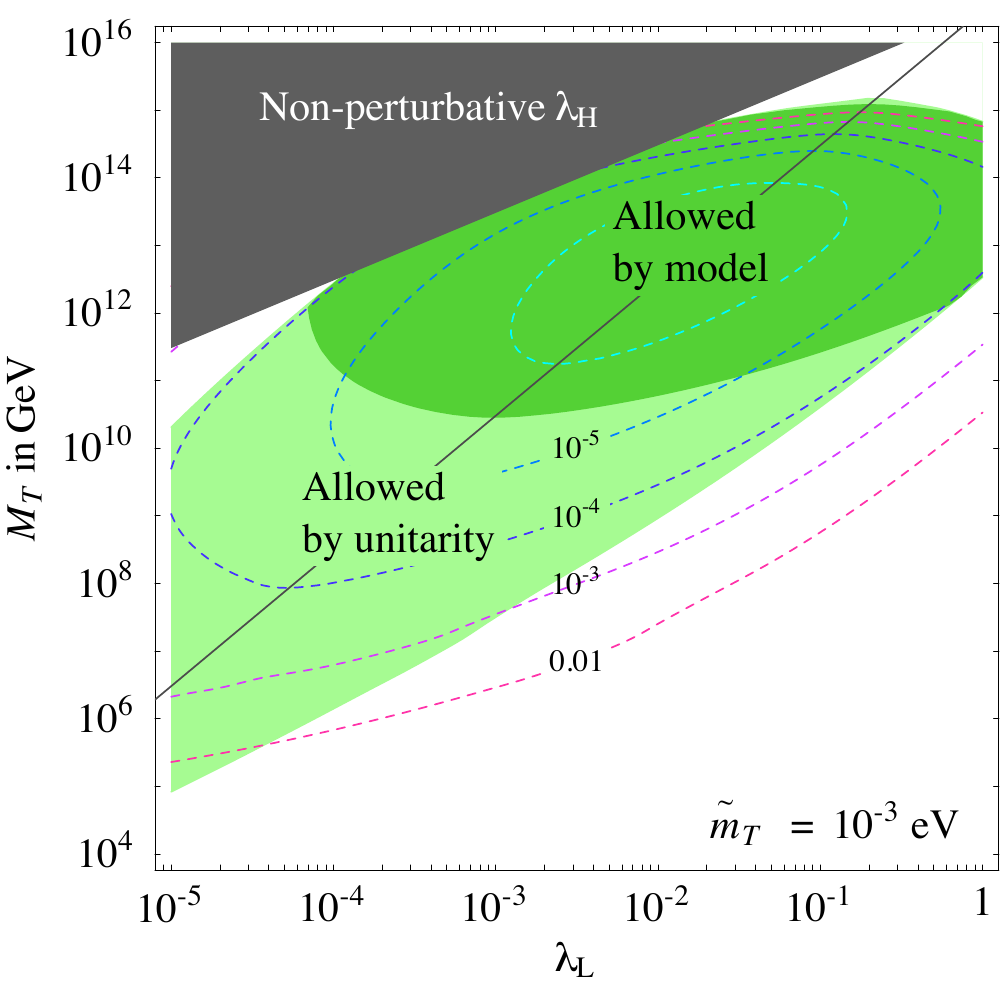}\includegraphics[height=7.5cm]{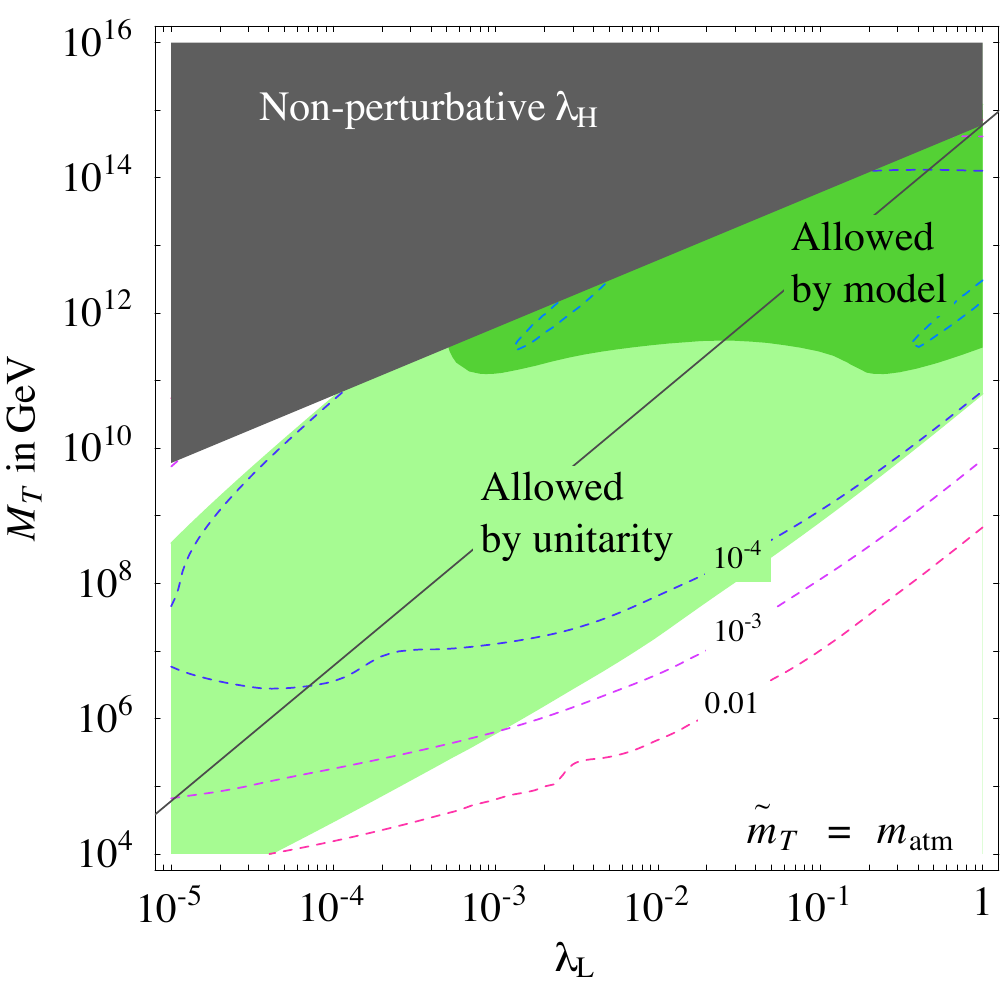}\\
\caption{From Ref.~\cite{Hambye:2005tk}, values of $\lambda_L$ and $m_\Delta$ which allows successful leptogenesis for a hierarchical spectrum (dark green, Eq.~(\ref{epsboundDelta2})) and from Eq.~(\ref{epsboundscalarunit}) (light green).
The dashed lines are isocurves of $\varepsilon_\Delta/\sqrt{4B_L B_H}$. The left (right) panel is obtained for $\tilde{m}_\Delta=10^{-3}$~eV ($=0.05$~eV). The shaded regions are regions where $\lambda_H\equiv \sqrt{2} \mu/m_\Delta$ is larger than one.}  
\label{etascal}
\end{figure}
Using the efficiency results above, Fig.~\ref{etascal} gives the region of $m_\Delta$ and $\lambda_L$ which can lead to successful baryogenesis, dark green region in Fig.~\ref{etascal}. To this end one has considered in Eq.~(\ref{nBoversscalar}) the bound on the CP-asymmetry which holds for a hierarchical seesaw state spectrum, Eq.~(\ref{epsboundDelta2}).
As one could expect, this CP asymmetry decreases off the $B_L=B_H$ case (i.e.~$\varepsilon_\Delta\propto \sqrt{B_LB_H}$), since in this case the amount of $L$ violation decreases.
Consequently the behavior of the CP-asymmetry is opposite to the one of the efficiency. 
Thus, one could wonder whether both effects compensate each other exactly in the baryon asymmetry produced, Eq.~(\ref{nBoversscalar}). In fact they do not cancel exactly. As explained above, in the 
regime where the decay term dominates the washout (as is the case for $\tilde{m}_\Delta=0.05$~eV and $m_\Delta \gtrsim 10^6$~GeV), the efficiency  goes approximately as $1/K_{min}\propto 1/Min[B_L,B_H]$. Therefore unless we are so far away from the $B_L=B_H=1$ case that the efficiency is unity (giving $n_B/s\propto \sqrt{B_L B_H}$), $n_B/s$ goes like $\sqrt{B_L B_H}/Min[B_L,B_H]$. As a result for fixed value of $m_\Delta$, the maximum baryon asymmetry is obtained off the $B_L=B_H=1/2$ case. For $\tilde{m}_\Delta=0.05$~eV, this can be seen in Fig.~\ref{etascal}. 
For  the $\tilde{m}_\Delta=10^{-3}$~eV case on the other hand there is no washout from the decays/inverse decays and the efficiency is independent of $B_{L,H}$. It depends only on $m_\Delta$ through the gauge scattering effect. Thus, the baryon asymmetry is proportional to $\sqrt{B_L B_H}$ in the same way as the CP-asymmetry and one gets a maximum baryon asymmetry for the $B_L=B_H=1/2$ case, see Fig.~\ref{etascal}. 

Since, for a hierarchical spectrum of seesaw states, the upper bound on the CP asymmetry, Eq.~(\ref{epsIIhierarc}), is proportional to $m_\Delta$, successful leptogenesis requires a heavy scalar triplet. Moreover the thermalization suppression effect from the gauge scattering increases as $1/m_\Delta$. Assuming a hierarchical spectrum of light neutrinos, this leads to the bounds \cite{Hambye:2005tk}
\begin{eqnarray}
m_\Delta&>& 2.8\cdot 10^{10}\,\hbox{GeV} \quad (\tilde{m}_\Delta=0.001~\hbox{eV})\,,\nonumber\\
m_\Delta&>& 1.3\cdot 10^{11}\,\hbox{GeV} \quad (\tilde{m}_\Delta=0.05~\hbox{eV})\,.
\label{deltamassbounds}
\end{eqnarray}
As for the fermion triplet case these bounds are stronger than for the type-I seesaw case due to the gauge scattering effects. These bounds apply using Eq.~(\ref{epsboundDelta}). For small contribution of the heavy seesaw states to the neutrino mass matrix a stronger bound applies. In this case the bounds of Eqs.~(\ref{deltamassbounds}) have to be multiplied by a factor $\sqrt{\sum m_{\nu_i}^2}/\tilde{m}_H$, see Eq.~(\ref{epsboundDelta2}). For example, 
for $\tilde{m}_H = (\Delta m^2_{\rm sun})^{1/2}=0.007$~eV this gives:
$m_\Delta  >  8 \cdot 10^{11}\,\hbox{GeV}$ (with $\tilde{m}_\Delta=0.05\,\hbox{eV}$).
These precise constraints have the same order of magnitude than the ones which were estimated in 
Refs.~\cite{Hambye:2000ui,Hambye:2003ka,Hambye:2004fn}.

For larger values of $\sum_i m_{\nu_i}^2$ the constraints of 
Eqs.~(\ref{deltamassbounds}) get relaxed by a factor
$(\Delta m^2_{\rm atm}/\sum_i m_{\nu_i}^2)^{1/2}$, i.e.~by  one order of magnitude for quasi-degenerate neutrinos with $m_\nu\approx0.5\hbox{eV}\approx 10 \,m_{\rm atm}$. This allows to go down to $m_\Delta\sim 10^9$~GeV.
In fact, by increasing the neutrino mass scale, keeping $m_\Delta$ fixed,
the asymmetry increases but the efficiency remains unchanged~\cite{Hambye:2003ka}.


\subsection{Leptogenesis possibilities with neutrino masses induced by a single scalar triplet}

The type-II seesaw model with a single scalar is the only seesaw model which can generate 
any possible neutrino mass matrix on the basis of a single state beyond the SM.
As a result, in full generality, this model involves "only" 11 parameters, the 
triplet mass, its $\mu_\Delta$ coupling and 6 real parameters plus 
3 phases 
in the $Y_\Delta$ Yukawa coupling matrix, see Table.~1. Out of these parameters, 3 of them are normalization parameters ($m_\Delta$, $\mu_\Delta$ and $y_\Delta=(Tr[Y_\Delta Y_\Delta^\dagger])^{1/2}$) and 8 are "flavour" parameters. This is similar to what one has with 2 type-I or III states, with the important difference that there are less decoupling parameters in the type-II option.
In particular this model has the attractive property that the knowledge of 
the full low energy neutrino mass matrix $M_\nu$ 
would allow to determine the full 
flavour high energy structure in $Y_\Delta$, both matrices being just 
proportional to each other. In this sense it has a "Minimal flavour structure", see e.g.~Ref.~\cite{Gavela:2009cd}, that is to say that the knowledge of the flavour structure of the lowest dimension operator (i.e.~the dimension-4 $Y_\Delta$ interaction), allows to know the flavour structure of all higher dimensional effects (i.e.~of the L-violating dim-5 neutrino mass matrix, of the seesaw induced L-conserving dim-6 operators generating rare lepton flavour violating processes such as $\mu\rightarrow e \gamma$, etc). This is not the case with 2 right-handed neutrinos, except in very particular cases \cite{Gavela:2009cd}.
As for the 3 normalization parameters, the neutrino mass matrix allows to know their combination $y_\Delta \mu_\Delta /m_\Delta^2$. Only 2 parameters decouple. 
This model would therefore constitutes a perfect minimal leptogenesis model, with all CP-violating phases which could be determined at low energy, from the neutrino mass matrix. However, unfortunately, it is so minimal that it doesn't work at all for leptogenesis, 
as already mentioned above and explained in Ref.~\cite{Ma:1998dx,Hambye:2000ui,Hambye:2003rt,D'Ambrosio:2004fz}.
Since the triplet is not 
a self-conjugated 
particle, there is no vertex diagram and leptogenesis could 
come only from a self-energy diagram involving two leptons in the final state
and two Higgs doublets in the self-energy, first diagram of Fig.~\ref{fig33}. 
This diagram, with just one triplet,
is real and doesn't bring any CP-violation. Flavour effects do not help on this issue. And at the higher loop level the 
asymmetries would be too suppressed anyway (and this model involves only one source of flavour breaking anyway too).
Therefore the  standard model 
extended by just one scalar triplet leads in a ``minimal'' 
way to neutrino masses but do not lead to 
successful leptogenesis.

There are nevertheless possibilities to have successful leptogenesis with neutrino masses induced only by a single triplet.
One possibility arises within the supersymmetric version of this single triplet model. This model involves an additional triplet of opposite hypercharge (to insure that the theory is anomaly free) but still only one scalar triplet contributing to the neutrino mass matrix. In this case successful leptogenesis is possible from a self energy diagram involving both triplets and extra soft susy breaking sources of $L$ violation \cite{D'Ambrosio:2004fz,Chun:2005ms}. The CP asymmetry in this case is determined by both the flavour structure of the $Y_\Delta$ coupling and of the L violating soft terms. To have a non-vanishing CP asymmetry in this framework, it is sufficient that the various superpotential terms and soft terms differ only by an overal non-vanishing phase and that both triplets mix and have a non-vanishing mass splitting due to soft terms (bilinear in the triplets).

Another interesting possibility, which doesn't rely on soft susy breaking terms, arises in a SO(10) model based on a particular set of representations \cite{Frigerio:2008ai}. It consists in assuming, in addition to the usual $10_H$ and $16_{F_{1,2,3}}$ SO(10) representations, $10_{F_{1,2,3}}$ fermions representations, a $54_S$ scalar representation and a $10_S$ scalar representation. The ordinary SM lepton doublets are in the $10_F$ multiplets. As a result there is no type-I contribution to neutrino masses (i.e.~the usual $Y_N 16_F 16_F 10_H$ do not involve the left-handed neutrinos) but a pure type-II one from the exchange of the scalar triplet in the $54_S$ between a pair of Higgs doublets in the $10_H$ and a pair of lepton doublets (i.e.~a $f_{ij}54_S10_{F_i}10_{F_j}$ coupling). In the supersymmetric version of this model leptogenesis is induced exclusively by an asymmetry involving 4 powers of the $f_{ij}$ couplings (through a vertex type diagram where the triplet decays to a pair of leptons (sleptons) with a virtual particle from the $54_S$). The second source of flavour breaking, required to have a CP asymmetry,  is provided by the mass hierarchy between the various beyond the standard model particles in the $10_{F_i}$ representations.
If we take for example the 2 heaviest ones to be heavier than $m_\Delta$ there is no GIM cancellation and we get an asymmetry proportional to $Im[f_{11}(f^*ff^*)_{11}]$ (where "1" refers to the lightest $10_{F_i}$ particle) times a function of the masses of the various particles. That's basically the maximum connection one could get between the neutrino masses and leptogenesis: a CP asymmetry totally determined by the neutrino mass matrix (here proportional to four power of the neutrino mass matrix with, in particular, CP-violation determined by it) times an unknown normalization constant.  

\subsection{Quasi-degenerate case: the double type-II seesaw model}

If there is more than one heavy scalar triplet contributing to the neutrino masses, leptogenesis can be 
easily induced by the decay of the triplets to two leptons with
a one-loop self-energy diagram involving two different 
triplets \cite{Ma:1998dx,Hambye:2000ui},
first diagram of  Fig.~\ref{fig33}. This is actually the first model which has been proposed 
to induce leptogenesis from the decay of a scalar triplet.
The asymmetry in this case is given by
\begin{equation}
\varepsilon_{\Delta_i} =  -\frac{1}{2 \pi} m_{\Delta_i}
\frac{ Im[(\mu^*_i \mu_j (Y_{\Delta_i})_{kl}
 (Y^*_{\Delta_j})_{kl} ] }{  |(Y_{\Delta_i})_{kl}|^2 m^2_{\Delta_i} 
+|\mu_i|^2}  \frac{M^2_{\Delta_j}  
\Delta m^2_{ij}}{(\Delta m^2_{ij})^2+m_{\Delta_i} ^2
   \Gamma_{\Delta_j} ^2}
\,,
\label{epsDi}
\end{equation}
where there is now a triplet scalar index on the couplings 
of Table.~1, with $m^2_{ij}=m^2_{\Delta_j}-m^2_{\Delta_i}$. 
For hierarchical triplets, all the results above hold. In particular the CP asymmetry of Eq.~(\ref{epsDi})
reduces to the one of Eq.~(\ref{epsIIhierarc}) for $m_{\Delta_2}>>m_{\Delta_1}$.
Similarly the bounds of Eq.~(\ref{deltamassbounds}) hold.
For quasi-degenerate triplets, however, a resonance effect occurs in Eq.~(\ref{epsDi}), similarly to Eq.~(\ref{epsIIIresonant}). This effect can lead to a large enhancement of the CP-asymmetry. As a result successful leptogenesis can be obtained for much smaller values of the triplet masses. Applying the bound of Eq.~(\ref{epsboundscalarunit}), one gets the absolute bound \cite{Strumia:2008cf}
\begin{equation}
m_\Delta> 1.6\,\hbox{TeV}\,.
\label{mdeltaaa}
\end{equation}
Similarly to what happens in the fermion triplet quasi-degenerate case above, this scale results from the interplay of the electroweak scale, below which 
there is no more $L\rightarrow B$ conversion, and of the gauge scatterings (which at those scales give a suppression of the efficiency of more than 5 orders of magnitudes, see Fig.~\ref{etastrumiascal}).
The value of Eq.~(\ref{mdeltaaa}) is too high to be reachable at LHC. Similarly to the fermion triplet case too, the LHC experiments should be able to discover a scalar triplet with a mass up to $\sim$~TeV \cite{TypeIILHC}-\cite{TypeIILHC6} (through Drell-Yan pair production). As of today the premiminary LHC lower bound on a doubly charged scalar boson is $\sim400$~GeV.

%

Note also that with several triplets, and whatever is the triplet mass spectrum,
there is no relevant upper bound on neutrino masses for successful 
leptogenesis because one can always increase the neutrino mass contribution
of the virtual triplet in the self-energy diagram, leading to a 
larger CP asymmetry, without affecting the efficiency.
For specific leptogenesis models involving several scalar triplets see Refs.~\cite{Chen:2010uc,Arina:2011cu}.

The supersymmetric version of this double type-II scenario has been considered in Ref.~\cite{Hambye:2000ui}. It involves two pairs of opposite hypercharge triplets (in order that the model is anomaly free). It differs from the non-supersymmetric version by factors of order unity for the CP-asymmetry and efficiency. In particular, for a hierarchical spectrum of triplet pairs, the bound of Eq.~(\ref{deltamassbounds}) remains valid and lies above the gravitino upper bound on the reheating temperature. Successful leptogenesis would therefore require in this framework a quasi-degenerate spectrum of triplet pairs, or extra L-violating soft susy breaking terms as proposed in Refs.~\cite{D'Ambrosio:2004fz,Chun:2005ms}.

\subsection{Effect of flavour}

Finally note that effects of flavour on the production of a lepton asymmetry from the decay of a scalar triplet are in general expected to be somewhat smaller than for the right-handed neutrino, if one assumes that this triplet dominates the neutrino masses, simply because in this case a single seesaw state gives several neutrino masses. But, still, one could have interplays of flavour asymmetries similar to the ones one can have with right-handed neutrinos. This has been studied for a specific model in Ref.~\cite{JosseMichaux:2008ix}. Note that in the same way as for a fermion triplet above, it is interesting to point out that, by invoking flavour effects (reducing the washout from the inverse decays) one could go below the upper bounds of Eq.~(\ref{deltamassbounds}). But flavour effects do not allow to go below the bound of Eq.~(\ref{mdeltaaa}) because this bound is obtained deeply in the gauge regime  where flavour effects are negligible (along the same way as for the fermion triplet above, 
Eq.~(\ref{msigmaboundtev})).

\section{Mixed seesaw models}

\subsection{The type-I + type-II model: decay of the scalar triplet}

In the left right models \cite{Pati:1974yy,Mohapatra:1974hk,Senjanovic:1975rk,Senjanovic:1978ev,Mohapatra:1980yp} spontaneous braking of parity finds its origin in the vev, "$v_R$", of a "$\Delta_R$"  hypercharge 2 scalar boson, which is a triplet of the $SU(2)_R$ gauge symmetry. This vev induces the right-handed neutrino Majorana masses.  The parity partner of $\Delta_R$ is nothing but the type-II seesaw $\Delta_L$ triplet. Similarly in $SO(10)$ models which give right-handed neutrino masses in a renormalizable way, one also finds a type-II seesaw field on top of the right-handed neutrinos (in the $126_S$ multiplet which contains both the $\Delta_L$ and $\Delta_R$ fields).
In all these cases the relevant Lagrangian for what concerns the neutrino masses is just the sum of the type-I and II Lagrangians 
of Table 1.
To discuss this possibility it is necessary to consider two cases, 
depending on which particle is the lightest one, the scalar triplet 
$\Delta_L$ 
or the lightest right-handed neutrino $N_1$ \cite{Hambye:2003ka}.

If the scalar triplet is the lightest seesaw state, its decay naturally dominates the production of the asymmetry from a diagram of vertex type with a virtual right-handed neutrino,
second diagram of Fig.~\ref{fig33}, first displayed 
in Ref.\cite{O'Donnell:1993am,Hambye:2003ka}.
Calculating explicitly its contribution we get \cite{Hambye:2003ka,Hambye:2005tk}:
\begin{equation}
\varepsilon_\Delta =  -\frac{1}{16 \pi^2}
\frac{m_{N_k}^2 m_\Delta}{ \Gamma_\Delta}
\frac{1}{v^4}
Im[({\cal M}_\nu^\Delta)_{\beta \alpha
}({\cal M}^{I}_\nu)^\dagger_{\alpha\beta} ] \log[1+\frac{m_\Delta^2}{m_{N_k}^2}]
\,.
\label{epsD}
\end{equation}
For $m_{N_{1,2,3}}>>m_\Delta$ this asymmetry reduces to the one of Eq.~(\ref{epsIIhierarc}) with ${\cal M}_\nu^H=\sum_i{\cal M}_\nu^{N_i}$, and the bounds of Eq.~(\ref{deltamassbounds}) hold.
In this case there is obviously no relevant bounds on the neutrino masses \cite{Hambye:2003ka} because the asymmetry is quadratic in couplings which do not induce any washout, i.e.~in the type-I Yukawa couplings. This feature also allows to reduce the lower bound on $m_\Delta$ of Eq.~(\ref{deltamassbounds}), because if one increases these couplings, keeping fixed the scalar triplet neutrino mass contribution, one increases the CP-asymmetry without changing the efficiency. In this case, from Eq.~(\ref{epsboundDelta2}), the bounds of Eq.~(\ref{deltamassbounds}) are relaxed by a factor $0.05~\hbox{eV}/{(\hbox{Min}[\sum m_{\nu_i}^2,\sum m_{\nu_i}^{\Delta^2}])^{1/2}}$. For neutrino masses below the eV scale this doesn't allow to go below $\sim 10^{10}$~GeV. One could eventually go a few times lower by invoking additional flavour effects in the same way as for the fermion triplet above.
Since the diagram of Fig.~\ref{fig33} doesn't display any resonant feature, to take $m_\Delta\sim m_{N_1,2,3}$ doesn't give any enhancement and the bounds above remain valid also for this case, up to factors of order unity.

\subsection{The type-I + type-II model: decay of a right-handed neutrino}

If the lightest right-handed neutrino mass is lighter than the scalar triplet, one must consider the opposite case where the asymmetry production is dominated by the decay of the lightest right-handed neutrino. In this case leptogenesis can be produced from the pure type-I usual diagrams.\footnote{The contribution of the usual pure type-I CP-asymmetry in the context of type I + II seesaw models has been studied in details in Ref.~\cite{Joshipura:2001ui}-\cite{Joshipura4}.} However there is an additional contribution to leptogenesis which comes from the third diagram of Fig.~\ref{fig33}, involving a virtual scalar triplet \cite{O'Donnell:1993am,Lazarides:1998iq,Chun:2000dr,Hambye:2003ka,Antusch:2004xy}. This diagram gives an extra contribution to the CP asymmetry of the lightest right-handed neutrino \cite{Hambye:2003ka,Antusch:2004xy}
\begin{equation}
\hspace*{-1.3cm}
\varepsilon_{N_1}=-\frac{3}{8\pi^2}
\frac{m_{N_1}m^2_\Delta}{ \Gamma_{N_1}}
\frac{1}{v^4}
\hbox{Im}[(M_\nu^{N_1})_{\beta \alpha
}(M^{\Delta}_\nu)^\dagger_{\alpha\beta} ] \Big(1-\frac{m_\Delta^2}{m_{N_1}^2}
 \log[1+m_{N_1}^2/m_\Delta^2]\Big)\,,
\label{epsNscal}
\end{equation} 
which, in the hierarchical limit $m_{N_1}<<< m_\Delta$, reduces to Eq.~(\ref{epsIhierarc}). In fact in this hierarchical limit, and with $m_{N_1}<<<m_{N_{2,3}}$, the respective contribution of $N_{2,3}$ and $\Delta_L$ is proportionnal to their contribution to the neutrino masses. If the neutrino mass matrix is dominated by the scalar triplet, the scalar triplet contribution of Eq.~(\ref{epsNscal}) will naturally dominate the production of the baryon asymmetry. If instead $N_{2,3}$ gives a larger contribution to the neutrino mass matrix, the asymmetry of Eq.~(\ref{epsIhierarc}) with ${\cal M}^{H}_\nu$ replaced by ${\cal M}^{N_{2,3}}_\nu$ will naturally dominate. Of course, since we are dealing with matrices, this depends on the interplay of the various complex flavour entries in ${\cal M}_\nu^I$ and in ${\cal M}_\nu^{\Delta}$ or ${\cal M}_\nu^{N_{2,3}}$. Consequently there is no one to one correspondance between the size of neutrino masses induced by each seesaw state and their contribution to the CP asymmetry, but this is the natural pattern which emerges \cite{Hambye:2003ka}.
For instance the scalar triplet term of Eq.~(\ref{epsNscal}) could perfectly do the job alone.
This contribution, here too, doesn't decrease with neutrino masses, since it is quadratic in the scalar triplet couplings which do not cause any washout.
As a result, here too, there is no relevant upper bound on neutrino masses, even though the seesaw state have a hierarchical spectrum. Moreover given the fact that the CP-asymmetry is linear in $M_{N_1}$ this allows to decrease the bound on the lightest right-handed neutrino mass linearly in the neutrino masses \cite{Hambye:2003ka,Antusch:2004xy}. The precise bound one gets in this case on the CP-asymmetry is \cite{Antusch:2004xy}
\begin{equation}
\varepsilon_{N_1}<\frac{3}{8\pi}\frac{M_{N_1}}{v^2}m_\nu^{max}\,,
\end{equation}
with $m_\nu^{max}$ the highest neutrino mass eigenvalue. This gives
\begin{equation}
M_{N_1}> 6.0\cdot 10^8\,\hbox{GeV}\cdot\Big(\frac{0.05\,\hbox{eV}}{m_\nu^{max}}\Big)\,.
\end{equation}
For neutrino masses below the eV scale this allows to go down to $\sim \hbox{few}\,10^7$~GeV. Here too since the associated one-loop diagram is non-resonant one cannot go below this value. 
For various studies of the type-I +II leptogenesis frameworks, including within the context of specific flavour models, see Refs.~\cite{typeIIleptononGUT,typeIIleptononGUT2,typeIIleptononGUT3,typeIIleptononGUT4,typeIIleptononGUT5,typeIIleptononGUT6,typeIIleptononGUT7,typeIIleptononGUT8,typeIIleptononGUT9,Chun:2006sp,Branco:2011zb,LRlepto,LRlepto2,LRlepto3,LRlepto4,LRlepto5,LRlepto6,SO10lepto,SO10lepto2,SO10lepto3,SU5triplet,AristizabalSierra:2012js}.
A study of flavour effects in the type-I+type-II model producing the asymmetry from the right-handed neutrino decays (i.e.~with $m_{N_1}<< m_\Delta$) can be found in Ref.~\cite{Antusch:2007km}.

\subsection{Implications for left-right symmetric and SO(10) models:}

The most important prediction of the left-right models, for what concerns neutrino masses and leptogenesis, is that right and left-handed neutrino mass matrices are proportional to each other, i.e.~$(Y_\Delta)_{ij}=\frac{1}{2} \delta_{ij} m_{N_i}/v_R$, with $v_R$ the $SU(2)_R$ breaking scale. 
As a result from Eqs.~(\ref{epsD}) and (\ref{epsNscal}) one gets 
\begin{eqnarray}
\varepsilon_\Delta&=&-\frac{1}{4 \pi}\frac{m_{N_i}}{m_{N_k}}\frac{\hbox{Im}[ (Y_{N})^2_{ki} \mu_\Delta]}{\Big(\frac{1}{4}\frac{m_{N_j}^2}{v_R^2}+\frac{|\mu_\Delta|^2}{m^2_\Delta}\Big)}\,,\\
\varepsilon_{N_1}&=& \frac{3}{16\pi}\frac{m_{N_1}m_{N_i}}{m^2_\Delta v_R}\frac{\hbox{Im}[ (Y_{N})_{1i}^2 \mu_\Delta]}{ |Y_{N1j}|^2}\,,
\end{eqnarray}
which depends only on the normalization parameters $\mu_\Delta$, $m_\Delta$ and $v_R$, and on the right-handed neutrino parameters, $(Y_N)_{ki}$ and $m_{N_i}$. Thus, apart for normalization factors this framework involves in the CP-asymmetries the same number of parameters than the pure type-I framework.
This must be taken into account for quantitative analysis of the asymmetry, even if it has basically no effect on the above lower bounds on the mass of the decaying state.

In the left-right models the mass hierarchy of seesaw states is not predicted. One could have nevertheless the prejudice that the Yukawa couplings are hierarchical and that $m_\Delta$ gets contributions of order the $SU(2)_R$ breaking scale $v_R$ (provided the associated scalar coupling would be of order unity, for example the $\Delta_R^\dagger \Delta_R \Delta_L^\dagger \Delta_L$ coupling). This prejudice would point out towards the $m_{N_1}< m_\Delta$ scenario.
Note that in this case there is no real reasons to believe that the diagram involving the scalar triplet, Eq.~(\ref{epsNscal}), would contribute less than the pure type-I diagrams. On the contrary one could argue that to have a dominant type-II contribution to the neutrino masses is a particularly attractive way to obtain large neutrino mixing angles despite the fact that the quark mixings are small (especially in the SO(10) framework where quarks and lepton Yukawa couplings are related see e.g.~Ref.~\cite{Bajc:2002iw}).

Similarly in the left-right model the mass of the second Higgs doublet in the scalar bidoublet is not predicted. It is generally expected large as it receives contributions proportional to $v_R$ from the scalar potential, but nothing prevents it from being below $m_{N_1}$. In this case both scalar doublets appear in all diagrams, so that in the CP asymmetries the $Y_N$ couplings carry now a scalar index, $Y_N^{i}$, $i=1,2$, and $\mu_\Delta$ becomes a $2\times2$ matrix.  

Note also that in the left-right model the right-handed gauge bosons could also have an effect if their mass is similar to $m_{N_1}$. Since $m_{W_R}=g_R v_R/2\sim v_R$, and since the $SU(2)_R$ coupling $g_R$ is \`a priori of order one (or equal to the $SU(2)_L$ coupling in the left-right symmetric case), this is the case only if $m_{N_{1,2,3}}\sim v_R$. In this case the main effect of the $W_R$ and $Z_R$ is a washout effect \cite{Cosme:2004xs,Frere:2008ct}. This effect is especially dangerous for the efficiency because $W_R$ gauge interactions induce scattering processes involving only one external right-handed neutrino: $N e_R\rightarrow \bar{u}_R d_R$, $N \bar{u}_R\rightarrow \bar{e}_R {d}_R$, $N d_R\rightarrow {e}_R u_R$. Thus, unlike the gauge scatterings induced by left-handed gauge boson above, which involve two external seesaw states, these scatterings have a reaction rate $\gamma_A^R$ which is suppressed by only one Boltzmann suppression power instead of two. That means that they have a thermalization rate $\gamma_A^R/n_N^{eq}H$ which is not Boltzmann suppressed at all \cite{Frere:2008ct}! As a result if $m_{W_R}\sim m_{N_1}$ these scatterings maintain the $N_R$ in thermal equilibrium until $T<<  m_{N_1}$ where $n_N^{eq}$ is hugely suppressed. In this case the bulk of the asymmetry is produced (in a very suppressed way) before this decoupling. Only for $m_{N_1}\gtrsim 10^{14}$~GeV are these scatterings never in thermal equilibrium.
Consequently successful leptogenesis in this scenario basically requires $m_{N_1}<< m_{W_R}$ or $m_{N_1}\gtrsim10^{12}$~GeV   or $m_{\Delta_L}\lesssim m_{N_1}$. 
This has actually consequences for what would imply the discovery of a right-handed neutrino at the LHC. The right-handed neutrino production channel usually considered at LHC is through an intermediate $W_R$ produced from quarks. Such a process leads to a measurable signal only if the scale $m_{W_R}\sim v_R$ is not much above the mass of the right-handed neutrino, and doesn't 
exceed the $\sim 4$~TeV scale, see e.g.~Refs.~\cite{Ferrari:2000sp,cmsW}. The discovery of a right-handed neutrino and/or of a $W_R$ at LHC would therefore excludes the right-handed neutrino decay leptogenesis scenario \cite{Frere:2008ct}.  Successful leptogenesis requires $m_{W_R}$ above the $\sim 10$~TeV scale.  

For various leptogenesis studies in the frameworks of left-right models, see Refs.~\cite{LRlepto,LRlepto2,LRlepto3,LRlepto4,LRlepto5,LRlepto6,Frigerio:2006gx}.
Similarly for SO(10), see Refs.~\cite{SO10lepto,SO10lepto2,SO10lepto3,Antusch:2005tu}. In SO(10) models where the $Y_N$ couplings have the same flavour structure than the one of up quark mass matrix, the
type-I seesaw formula leads to a strong hierarchy between the $m_{N_i}$ (such that $m_{N_{1,2,3}}\propto m^2_{u,c,t}$). Given the GUT scale value, this gives a value of $m_{N_1}$ well below the bound of Eq.~(\ref{MNbound}). This scenario has been analyzed in details in the context of pure type-I SO(10) inspired models \cite{Branco:2002kt,Nezri:2000pb,Akhmedov:2003dg,SO10lepto3}. This problem can be cured in various ways, in particular if one has also a type-II contribution dominating the neutrino masses \cite{Antusch:2005tu}, or in the $N_2$-leptogenesis framework mentioned above, invoking flavour effects \cite{DiBari:2008mp}-\cite{DiBari3}.
For a related model based on $E_6$ grand-unification, see Ref.~\cite{Hambye:2000bn,King:2008qb}.

\subsection{The type-I+type-III model and implications for the adjoint SU(5) models}

As mentioned in section 2, a type-I + type-III neutrino mass generation setup is a quite straightforward possibility in the framework of grand-unification \cite{Bajc:2006ia,Dorsner:2006fx,Bajc:2007zf,Fischler:2008xm,Blanchet:2008cj,Kamenik:2009cb}. In the adjoint representation of $SU(5)$ there is a particle which has the quantum numbers of a seesaw fermion triplet and another particle which has the ones of a right handed neutrino. This leads to a particularly minimal model where all seesaw states are in a single representation. For what concerns neutrino masses it is similar in many respects to the 2 right-handed neutrinos seesaw models.
It can fulfill easily the neutrino mass data constraints. Moreover it may lead to gauge unification if the fermion triplet is light, below the $\sim2$-3~TeV scale. As mentioned above a triplet fermion with mass below $\sim$~TeV could be discovered at LHC. At such scale, for what concerns leptogenesis, one may think about the resonant self-energy possibility \cite{Bajc:2007zf}, especially for the decay of the right-handed neutrino whose efficiency is not suppressed by any gauge scatterings. That would require that both fermions have masses very close to each other. Beyond the fact that in this model there is no reason for such a pattern, this anyway doesn't work because such a self-energy diagram requires that a singlet fermion turns into a triplet fermion at the one loop level, which requires $SU(2)_L$ breaking. This turns out to lead to a very suppressed asymmetry at temperatures above the electroweak scale \cite{Blanchet:2008cj}.

  

\section{Conclusion}

The seesaw framework allows to explain within a common framework two experimental facts that, at first sight, one would have expected to be disconnected, the neutrino masses and the baryon asymmetry of the Universe. It is important to determine the main seesaw scenarios along which this could be achieved and to determine their main general properties.
It turns out that there exist quite a few attractive scenarios of this kind! Beside the usual right-handed neutrino decay scenario, the fermion triplet and scalar triplet decay scenarios could be perfectly work as well.  Each of these two scenarios can be realized in various ways, depending on the nature of the other seesaw states contributing to the neutrino masses. In these scenarios, as in the type-I framework, the neutrino masses have values within a range which turns out to be perfect for the generation of a large asymmetry.  An important difference, nevertheless, is that, if their mass is below $\sim 10^{14}$~GeV, these triplet states do have gauge interactions which thermalize them with the Universe thermal bath, at a temperature of order their masses. We devoted a large part of this review to explaining in details how the involved interplay of these interactions and the Yukawa interactions takes place in the calculation of the efficiency. Constraints  that this interplay do imply on the seesaw parameters are determined and compared with the type-I usual constraints.

An important question which arises is whether these attractive possible explanations of the baryon asymmetry of the Universe could be tested in a near future. There exist clear possibilities of falsifying the leptogenesis mechanism. For instance
it is well-known that a inverse hierarchical or quasi-degenerate spectrum of light neutrinos imply a lower bound on the size of the neutrinoless double beta decay signal \cite{Feruglio:2002af}.
If in the future one of these two mass spectra turns out to be established experimentally, the non-observation of such a process with such a minimum magnitude would basically rules out the seesaw mechanism as main origin of the neutrino masses. In a different vein the observation of a TeV scale right-handed $W_R$ boson would basically rules out the type-I leptogenesis, due to the huge washout of the L-asymmetry this state would imply. Similarly
a fermion or a scalar triplet that would be discovered at LHC, could easily wash-out any preexisting asymmetry. 
More generally the observation of new particles at LHC, other than seesaw particles, with lepton number violating interactions, could rule out leptogenesis scenarios.
As for establishing experimentally a leptogenesis scenario, this appears to be a tremendously difficult task. Triplet states, unlike right-handed neutrinos, do have the advantage that, from SM gauge interactions, they can be Drell-Yan pair produced at colliders. Unfortunately the efficiency suppression that gauge interactions imply at such low scales, leads, for successful leptogenesis, to a lower bound on the triplet state mass which is slightly above the reach of the LHC. In any case it must be kept in mind that even if at LHC one could observe a seesaw state with interactions which could produce the baryon asymmetry of the Universe, still this would not  preclude the possibility that the reheating temperature has been low enough to forbid this possibility, or that the baryon asymmetry which we observe today is the product of a dynamics which took place afterwards, for example along the supersymmetric Affleck-Dine mechanism (closely related to the dynamics of inflation). 
For all these reasons to establish the leptogenesis mechanism is even more difficult than to establish the seesaw mechanism as the origin of neutrino masses.
In this sense one could  say that leptogenesis remains more a fascinating paradigm -- that could well have been realized in Nature -- than a framework that one could test and study on and on with always greater accuracy.

\ack
I warmly thank D.~Aristizabal Sierra, S.~Blanchet, L.~Boubekeur, G.~D'Ambrosio, R.~Franceschini, J.-M. Fr\`ere, M.~Frigerio, F.-X.~Josse-Michaux, E.~Ma, M.~Papucci, M.~Raidal, A. Rossi, G.~Senjanovic, U.~Sarkar, A.~Strumia, M.~Tytgat and G.~Vertongen for many discussions and collaborations on leptogenesis.
This work is supported by the FNRS-FRS, the IISN and the Belgian Science Policy (IAP VI-11). 
Many thanks also to the Departamento de F\'isica Te\'orica (UAM-Madrid) and the IFT-Madrid for hospitality, and to the Comunidad de Madrid (Proyecto HEPHACOS S2009/ESP-1473).

\end{document}